\documentclass[11pt]{article}
\usepackage[T1]{fontenc}
\usepackage[utf8]{inputenc}
\usepackage{authblk}
\usepackage[american]{babel}
\usepackage{bm,amsfonts,amsmath,amssymb}
\usepackage{graphicx}
\usepackage[update]{epstopdf} 
\usepackage{subfigure}
\usepackage{epsfig}
\usepackage{color}
\usepackage{tikz,pgfplots}
\usepackage[a4paper]{geometry}
\usepackage{booktabs} 
\usepackage{tabulary,multirow}
\usepackage[normalem]{ulem}
\title{Impact of Airways Geometry on Transport of Gases to Blood}

\author[1]{Ali Saab}
\author[2]{Leila Issa}
\author[3]{Salah Zeineddine}
\author[4]{Daniel M. Tartakovsky}
\author[1]{Issam Lakkis}
\affil[1]{Department of Mechanical Engineering, American University of Beirut, Beirut, Lebanon}
\affil[2]{Department of Computer Science \& Mathematics, Lebanese American University, Beirut, Lebanon}
\affil[3]{Department of Internal Medicine, American University of Beirut, Beirut,  Lebanon}
\affil[4]{Department of Energy Resources Engineering, Stanford University, 367 Panama Street, Stanford, CA 94305, USA}

\date{}

\begin{document}


\maketitle

\begin{abstract}
Topological structure of bronchial trees affects transport of gases and aerosols in the respiratory system. We  start by providing a quantitative assessment of the alternative tree representations' ability to predict observable geometric and mechanistic characteristics, such as network resistance, dead space volume, and path length.  Then we present a model of dynamic transport of oxygen and carbon dioxide along the airways, in the alveoli, across the alveolar membrane, and along the pulmonary blood capillaries. The model also accounts for the exchange of these two gases with blood in the capillaries, as well as for age, gender and other in-species characteristics. Our model's predictions are compared with corresponding observations, providing an additional venue to assess the validity of the existing representations of the lung's bronchial tree. 
\end{abstract}

\section{Introduction}

Geometric structure of airways controls, to a large extent, transport of gases and aerosols in the respiratory system. These airways form a complex branching network, whose mathematical descriptions (e.g., connectivity, an airway's length-to-diameter ratio, etc.) underpin most quantitative studies in dosimetry~\cite{Miller-1993-Lower,menache2008airway} and in respiratory and obstructive diseases~\cite{Calay-2002-numerical,Smith-2018-Human}. Mathematical models of the bronchial tree in the human lung typically postulate a scaling law that relates the (generation-averaged) characteristics (diameter, length and branching angle) of each bronchial generation to those of the previous and successive generations.

There is a general consensus that the human bronchial tree forms a bifurcating network~\cite{Smith-2018-Human,Montesantos-2016-creation}. The nature of  bifurcations appears to be settled as well: while earlier and/or highly idealized models assume bifurcation symmetry~\cite{menache2008airway, weibel1963morphology, lee2007fluid}, more recent investigations allow bifurcations in bronchial trees to be asymmetric~\cite{Smith-2018-Human, Montesantos-2016-creation, Phillips-1997-on}. Finally, guided by morphometric data~\cite{raabe1976tracheobronchial, phalen1978application, weibel2014morphometry}, most (if not all) network models assume certain geometric properties scale across generations of the bronchial tree. Examples of such scaling laws are Weibel's~\cite{weibel1963morphology} and Horsfield's~\cite{horsfield1968morphology} models of bifurcation. The former posits that, in symmetrically bifurcating trees, a branch size $S$ (e.g., its length and diameter) scales with the branching order $n$ as $S(n) = S(0) \alpha^n$; in other words, the knowledge of the trachea size, $S(n=0)$, and the scaling factor $\alpha < 1$ is sufficient to estimate the branch size in any generation $n$ of the bronchial tree. The latter uses the same form of scaling law, $S(m) = S(0) \beta^m$, but groups the airways in order to account for the network asymmetry and starts the counting from the periphery upward, such that $S(m=0)$ is the smallest size and $\beta > 1$.

The parameters in these, and more evolved~\cite{west1986beyond, nelson1990fractal}, models are obtained by optimizing the bronchial tree's function. There is considerable debate about which specific cost function of the lung is to be optimized. It has been argued that the bronchial trees have evolved in a way that their bronchus' length-to-diameter ratio minimizes the total pressure loss across the bronchi~\cite{lee2007fluid}, that their asymmetrically bifurcating structure minimizes either dissipation of frictional energy~\cite{Thompson-1942-Growth} or a linear combination of the tree's resistance and volume~\cite{mauroy2010influence}, maximizes the volumetric flow rate~\cite{Olson1970}, etc. Natural in- and intra-species variability add a certain degree of randomness to the bronchial tree's structure~\cite{mauroy2010influence}, making it all but impossible to determine which cost function is ``correct''. 

To assess the veracity of alternative tree geometries, we compare predictions of mathematical models of air flow in the  bronchial networks with relevant observations. We start by providing a quantitative assessment of the existing network and gas flow models' ability to predict observable geometric and mechanistic characteristics, such as network resistance, dead space volume, and path length. The choice of these metrics is justified by the availability of their (relatively) high-fidelity experimental measurements. Then we present a model of dynamic transport of oxygen and carbon dioxide along the airways, in the alveoli, across the alveolar membrane, and along the pulmonary blood capillaries. The model also accounts for the exchange of these two gases with blood in the capillaries, as well as age, gender and other in-species characteristics. Our model's predictions are compared with corresponding observations, providing an additional means to assess the veracity of the existing bronchial tree geometries. We are not aware of any other comparison of the veracity of alternative bronchial tree models in terms of their ability to predict blood gas transport observables.

The novelty of our work is twofold: we (i) develop a novel model of gas transport in the lungs and (ii) use to build an an assessment tool, which builds on that model to evaluate alternative airways geometries found in the literature.  To highlight the innovation of our model of gas flow in the lung, we provide a brief overview of existing modeling strategies.

Models of gas transport in the lung can be subdivides into two categories: lumped models and spatially distributed models, which are typically formulated in terms of ordinary differential equations and partial differential equations, respectively.  Lumped models are derived from compartmental analysis and vary in complexity. In its original formulation~\cite{riley1951analysis}, a lumped model  divides the lung into three functional compartments: ideal, unventilated, and unperfused. This model serves as the starting point for the development of progressively more sophisticated models, such as VentPlan~\cite{rutledge1993design}, 
VenSim~\cite{rutledge1994ventsim}, 
and the Nottingham Physiology Simulator~\cite{hardman2001respiratory}. 
A hierarchy of lumped models of the human lungs, which describe the mechanical behavior of the lung (considered as a single compartment container) and gas exchange between the airways (another compartment) and the blood (the third compartment)
is proposed in~\cite{ben2006simplified}. These and other lumped models are computationally efficient, but have several limitations that preclude their use for the assessment of the alternative airway geometries. First, they do not predict a spatial distribution of gases in the airways and, hence, cannot be used to investigate the impact of airways geometry (e.g., dimensions of the branches along the tree) on gas exchange.
Second, they rely on fitting parameters that calibrated to match some selected quantities with measurements from patients; this precludes their use for assessment of the degree to which a given airways geometry is functionally similar to that of humans. 

Distributed models, including the one presented in this study, resolve the spatial variability in gas concentration along individual vessels forming the airways. 
Most, if not all, of such models use advection-reaction equations to model O$_2$ transport during the respiratory process~\cite{davidson1974transport, paiva1984model, swan2010evidence}, 
without accounting for the presence of CO$_2$. The most recent study of this kind~\cite{martin2013modeling} describes transfer of oxygen into the blood, enabling one to investigate the impact of longitudinal variability in O$_2$ concentration on the mean transfer rate. Our model extends the latter study by incorporating both transport of CO$_2$ and biochemical reactions between O$_2$ and CO$_2$ in the blood in the capillaries.

\section{Materials and Methods}

\subsection{Review of bronchial tree geometries}

\begin{figure}[htbp]
\centering
 \includegraphics[scale=0.6]{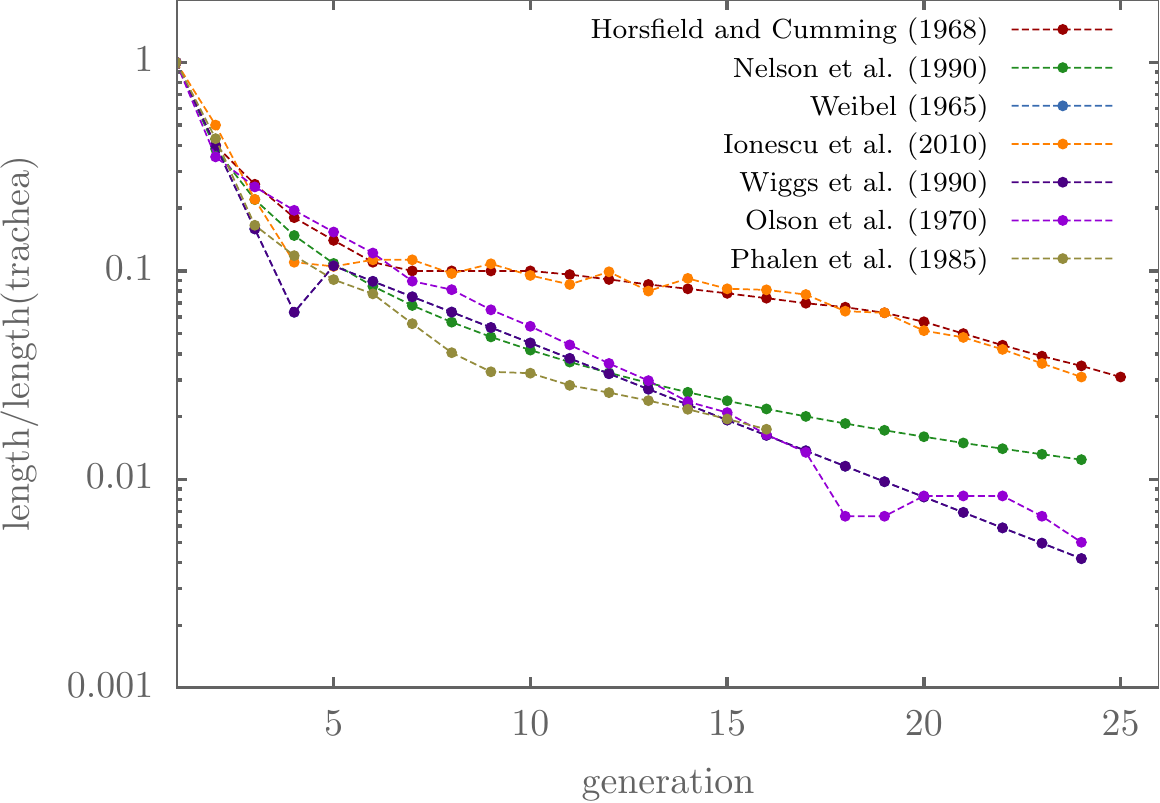}
 \includegraphics[scale=0.6]{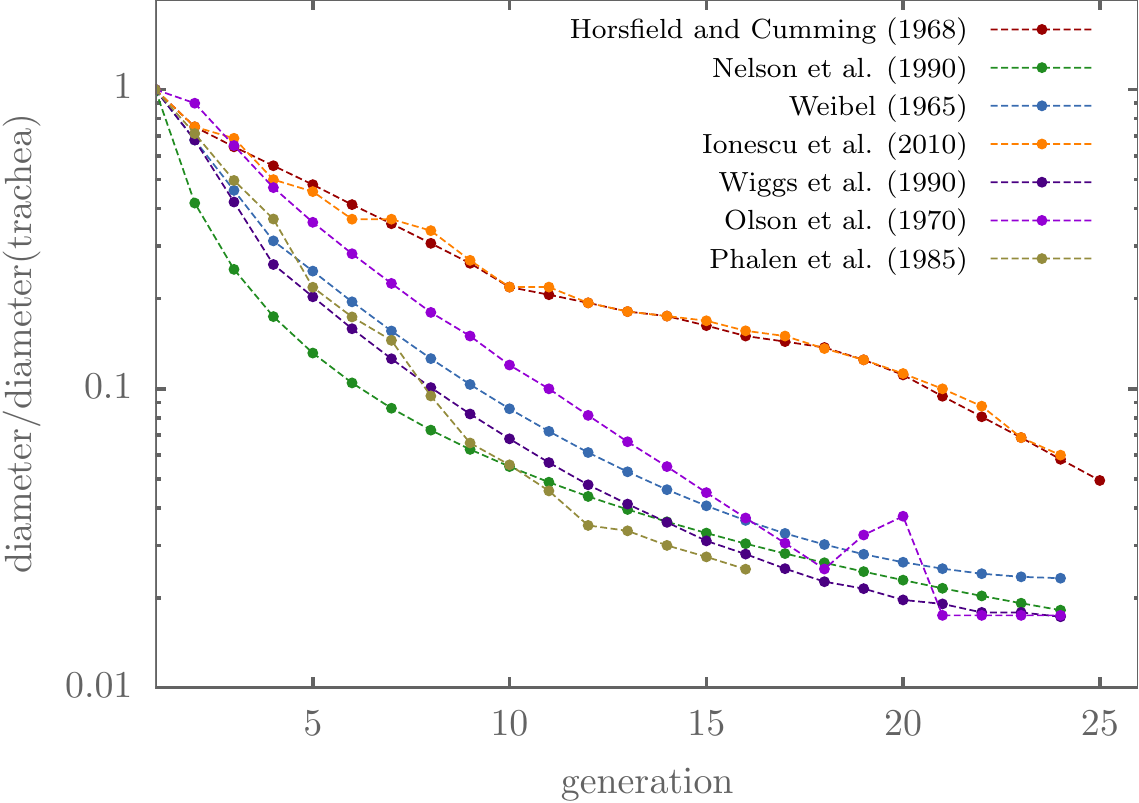}
 \includegraphics[scale=0.6]{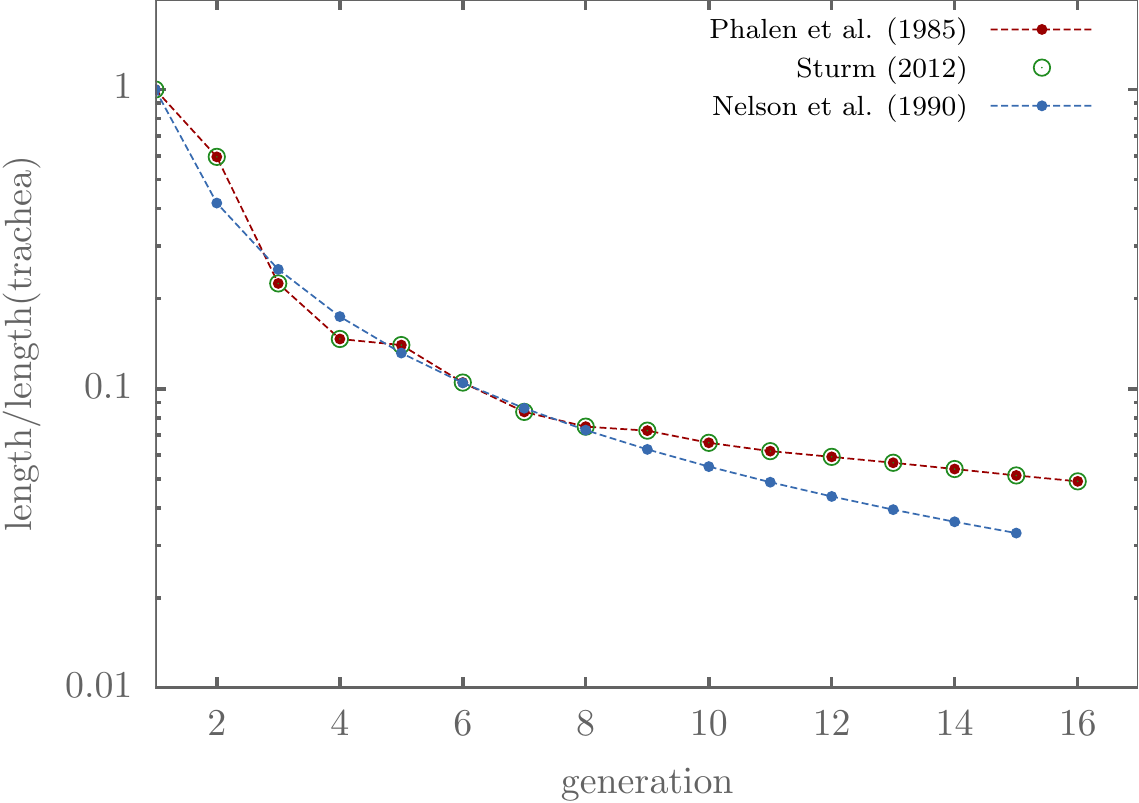}
 \includegraphics[scale=0.6]{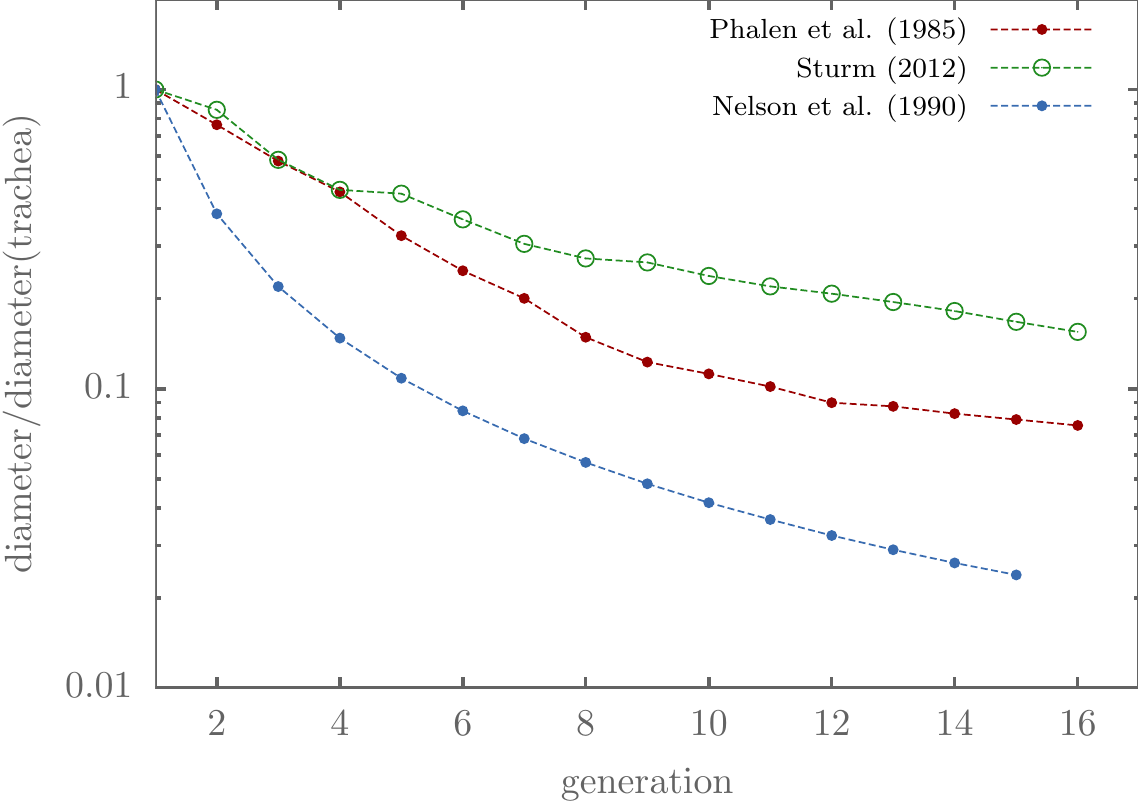}
\caption{Dependence of the length (left column) and diameter (right column), normalized with the corresponding length and diameter of the trachea, on the generation order $n$, in human adults (top row) and infants (bottom row).}
\label{figureAWLA}
\end{figure}

Available data on the (generation-averaged) length $L(n)$ and diameter $D(n)$ of individual bron\-chus in the $n$th generation ($n = 1,\cdots,24$) of the human bronchial tree are collated in Tables~\ref{Table:models2} and~\ref{Table:models3}, and presented graphically in Figure~\ref{figureAWLA}. These data are separated by age, such that Table~\ref{Table:models2} and the top row in Figure~\ref{figureAWLA} present data for human adults, while Table~\ref{Table:models3} and the bottom row in Figure~\ref{figureAWLA} are for human infants. These data served to inform multiple models of the human bronchial tree, six of which are described below.

\begin{table}[htbp]
\caption{Length ($L$) and diameter ($D$), both in mm, of the $n$th bronchi generation, predicted with tree models A--G for human adults. Model A~\cite{nelson1990fractal} infers $L \sim n^{-1.38}$ and $D \sim n^{-1.26}$ from data; Model B~\cite{weibel1965morphometry} $L \sim \exp(-0.92 n)$ and $D \sim \exp(-0.39 n)$ for $n \le 3$, and $L \sim \exp(-0.17 n)$ and $D \sim \exp(-0.29 n + 0.01 n^2)$ for $n > 3$; Model D~\cite{horsfield1968morphology} $D \sim a^{25-n}$ with $a > 1$ varying between three groups of bronchi characterized by three intervals of $n$; and Model E~\cite{phalen1985postnatal} identifying a linear relation, for each generation, between bronchi's length \& diameter and human height $H$, $L \sim H$ and $D \sim H$. The remaining tree models, C~\cite{ionescu2010viscoelasticity}, F~\cite{wiggs1990model} and G~\cite{Olson1970}, fit no curves to the data reported therein. }
\begin{center}
\vspace{0.1in}
{\scriptsize{
\begin{tabular}{ccccccccccccccc}
\hline
 & \multicolumn{2}{c}{A~\cite{nelson1990fractal} }&  \multicolumn{2}{c}{B~\cite{weibel1965morphometry} } & \multicolumn{2}{c}{C~\cite{ionescu2010viscoelasticity} } &  \multicolumn{2}{c}{D~\cite{horsfield1968morphology} } &  \multicolumn{2}{c}{E~\cite{phalen1985postnatal}  } & \multicolumn{2}{c}{F~\cite{wiggs1990model} }&\multicolumn{2}{c}{G~\cite{Olson1970} }\\
\cmidrule(lr){2-3}\cmidrule(lr){4-5}\cmidrule(lr){6-7}\cmidrule(lr){8-9}\cmidrule(lr){10-11}\cmidrule(lr){12-13}\cmidrule(lr){14-15}
\scriptsize{$n$} &$L$      &$D$&$L$ &$D$&$L$  &$D$&$L$  &$D$&$L$ &$D$&$L$ &$D$&$L$  &$D$\\
\midrule
1	&	120	&	18	&	120	&	18	&	100	&	16	&	100	&	16	&	89	&	18.2	&	120	&	16.8	&	120	&	20	\\
2	&	46.1	&	7.5	&	47.8	&	12.21	&	50	&	12	&	40	&	12	&	38.3	&	13.0	&	47.8	&	11.4	&	42.2	&	18	\\
3	&	26.4	&	4.5	&	19.1	&	8.3	&	22	&	11	&	26	&	10.3	&	14.7	&	9.0	&	19.1	&	7.1	&	30.3	&	13	\\
4	&	17.7	&	3.1	&	7.6	&	5.6	&	11	&	8	&	18	&	8.9	&	10.5	&	6.7	&	7.6	&	4.4	&	23.4	&	9.4	\\
5	&	13.0	&	2.4	&	12.7	&	4.5	&	10.5	&	7.3	&	14	&	7.7	&	8.1	&	4.0	&	12.7	&	3.4	&	18.4	&	7.2	\\
6	&	10.2	&	1.9	&	10.7	&	3.5	&	11.3	&	5.9	&	11	&	6.6	&	6.91	&	3.16	&	10.7	&	2.7	&	14.6	&	5.7	\\
7	&	8.2	&	1.6	&	9.0	&	2.8	&	11.3	&	5.9	&	10	&	5.7	&	5.0	&	2.6	&	9.0	&	2.1	&	10.7	&	4.5	\\
8	&	6.8	&	1.3	&	7.6	&	2.3	&	9.7	&	5.4	&	10	&	4.9	&	3.6	&	1.7	&	7.6	&	1.7	&	9.8	&	3.6	\\
9	&	5.8	&	1.1	&	6.4	&	1.9	&	10.8	&	4.3	&	10	&	4.2	&	2.9	&	1.2	&	6.4	&	1.4	&	7.8	&	3	\\
10	&	5	&	1.0	&	5.4	&	1.5	&	9.5	&	3.5	&	10	&	3.5	&	2.9	&	1.0	&	5.4	&	1.1	&	6.5	&	2.4	\\
11	&	4.4	&	0.9	&	4.6	&	1.3	&	8.6	&	3.5	&	9.6	&	3.3	&	2.5	&	0.8	&	4.6	&	1.0	&	5.3	&	2	\\
12	&	3.9	&	0.8	&	3.9	&	1.1	&	9.9	&	3.1	&	9.1	&	3.1	&	2.3	&	0.6	&	3.9	&	0.8	&	4.3	&	1.6	\\
13	&	3.5	&	0.7	&	3.3	&	1.0	&	8	&	2.9	&	8.6	&	2.9	&	2.1	&	0.6	&	3.3	&	0.7	&	3.6	&	1.3	\\
14	&	3.1	&	0.7	&	2.7	&	0.8	&	9.2	&	2.8	&	8.2	&	2.8	&	1.9	&	0.6	&	2.7	&	0.6	&	2.8	&	1.1	\\
15	&	2.9	&	0.6	&	2.3	&	0.7	&	8.2	&	2.7	&	7.8	&	2.6	&	1.7	&	0.5	&	2.3	&	0.5	&	2.5	&	0.9	\\
16	&	2.6	&	0.6	&	2.0	&	0.7	&	8.1	&	2.5	&	7.4	&	2.4	&	1.6	&	0.5	&	2.0	&	0.5	&	2.0	&	0.7	\\
17	&	2.4	&	0.5	&	1.7	&	0.6	&	7.7	&	2.4	&	7	&	2.3	&		&		&	1.7	&	0.4	&	1.6	&	0.6	\\
18	&	2.2	&	0.5	&	1.4	&	0.5	&	6.4	&	2.2	&	6.7	&	2.2	&		&		&	1.4	&	0.4	&	0.8	&	0.5	\\
19	&	2.1	&	0.4	&	1.2	&	0.5	&	6.3	&	2	&	6.3	&	2	&		&		&	1.2	&	0.4	&	0.8	&	0.7	\\
20	&	1.9	&	0.4	&	1.0	&	0.5	&	5.2	&	1.8	&	5.7	&	1.8	&		&		&	1.0	&	0.3	&	1	&	0.8	\\
21	&	1.8	&	0.4	&	0.8	&	0.5	&	4.8	&	1.6	&	5	&	1.5	&		&		&	0.8	&	0.3	&	1	&	0.4	\\
22	&	1.7	&	0.4	&	0.7	&	0.4	&	4.2	&	1.4	&	4.4	&	1.3	&		&		&	0.7	&	0.3	&	1	&	0.4	\\
23	&	1.6	&	0.4	&	0.6	&	0.4	&	3.6	&	1.1	&	3.9	&	1.1	&		&		&	0.6	&	0.3	&	0.8	&	0.4	\\
24	&	1.5	&	0.3	&	0.5	&	0.4	&	3.1	&	1.0	&	3.5	&	0.9	&		&		&	0.5	&	0.3	&	0.6	&	0.4	\\
25	&		&		&		&		&		&		&	3.1	&	0.8	&		&		&		&		&		&
\\ \hline
\end{tabular}
}}
\end{center}
  \label{Table:models2}
\end{table}

\textit{Geometry A}. The fractal lung model~\cite{west1986beyond, nelson1990fractal} assigns a probabilistic density function to   scaling factors for the dimensions of parent and daughter bronchi and employs the renormalization group theory to account for  the heterogeneity in the asymmetric airway branching. 
The model predicts the average dimensions (diameter, length, and volume) of the airways to obey power laws in the generation number $n$ (see the caption of Table~\ref{Table:models2}), with the model parameters inferred by fitting the model to the data from~\cite{weibel1963morphology, raabe1976tracheobronchial}. 

\textit{Geometry B.} The symmetrically bifurcating model of Weibel~\cite{weibel1963morphology, weibel2014morphometry} predicts the diameter and length of a bronchus to follow an exponential law in the generation number $n$ (see the caption of Table~\ref{Table:models2}). The model was inferred from, and parameterized by, a complete set of measurements of the first six generations ($1\le n \le 7$) of airways' diameters and lengths, some measurements from generations $8\le n \le 11$, and morphometric estimates of the lengths and diameters of respiratory bronchioles and alveolar airways ($n \ge 12$). 

\textit{Geometry C}. This model also assumes a fractal structure of the branching, with a self-similarity distribution that is preserved up to the 24th generation. The dimensions of the airways are inferred from prior observations~\cite{gheorghiu2005lung, harper2001acoustic, mauroy20053d, weibel2005mandelbrot, lee2007fluid, weibel1963morphology}.

\textit{Geometry D}.  
The asymmetric bifurcation model~\cite{horsfield1968morphology, horsfield1971models} orders airways by their relation to the periphery, as opposed to the trachea,  groups them by similar size, and subdivides them  into zones according to their degree of asymmetry. 
This ensures that the airways of the same order are more uniform in size than the airways of the same generation in model B~\cite{weibel1963morphology, weibel2014morphometry}.  Measurements of the complete human bronchial tree~\cite{horsfield1968morphology, horsfield1971models} are used to linearly correlate bronchi length $L$ with their diameter $D$.

\begin{table}[htbp]
\centering
\caption{Length ($L$) and diameter ($D$), both in mm, of the $n$th bronchi generation, predicted with the tree geometries A~\cite{nelson1990fractal}, E~\cite{phalen1985postnatal}, and H~\cite{sturm2012theoretical} for human infants.}
\begin{center}
\vspace{0.1in}
\begin{tabular}{ccccccc}
\hline
 & \multicolumn{2}{c}{A~\cite{nelson1990fractal} } &  \multicolumn{2}{c}{E~\cite{phalen1985postnatal} }  & \multicolumn{2}{c}{H~\cite{sturm2012theoretical} } \\
\cmidrule(lr){2-3}\cmidrule(lr){4-5}\cmidrule(lr){6-7}
$n$ & $L$      & $D$ & $L$ &$D$&$L$  &$D$\\
\midrule
1	&	26.5	&	5.69	&	26.5	&	5.69	&	26.5	&	3.28	\\
2	&	11.06	&	2.19	&	15.79	&	4.34	&	15.79	&	2.8	\\
3	&	6.64	&	1.25	&	5.96	&	3.28	&	5.96	&	1.91	\\
4	&	4.62	&	0.84	&	3.89	&	2.59	&	3.89	&	1.51	\\
5	&	3.49	&	0.62	&	3.71	&	1.85	&	3.71	&	1.47	\\
6	&	2.77	&	0.48	&	2.78	&	1.41	&	2.78	&	1.21	\\
7	&	2.28	&	0.39	&	2.22	&	1.14	&	2.22	&	1	\\
8	&	1.93	&	0.32	&	1.98	&	0.85	&	1.98	&	0.89	\\
9	&	1.66	&	0.27	&	1.92	&	0.7	&	1.92	&	0.87	\\
10	&	1.46	&	0.24	&	1.75	&	0.64	&	1.75	&	0.78	\\
11	&	1.29	&	0.21	&	1.64	&	0.58	&	1.64	&	0.72	\\
12	&	1.16	&	0.18	&	1.57	&	0.51	&	1.57	&	0.68	\\
13	&	1.05	&	0.17	&	1.5	&	0.5	&	1.5	&	0.64	\\
14	&	0.95	&	0.15	&	1.43	&	0.47	&	1.43	&	0.6	\\
15	&	0.87	&	0.14	&	1.36	&	0.45	&	1.36	&	0.55	\\
16	&		&		&	1.3	&	0.43	&	1.3	&	0.51 \\
\hline
\end{tabular}
\end{center}
\label{Table:models3}
\end{table}

\textit{Geometry E}. The model~\cite{phalen1985postnatal} accounts for the bronchi orientation to gravity and angles at branching points, and establishes linear relationships between a bronchus' length/diameter and the body height. This information is absent in both symmetric~\cite{weibel1963morphology, weibel2014morphometry} and asymmetric~\cite{horsfield1968morphology, horsfield1971models} models.The model is derived from a set of observations (casts) collected in humans, dogs, rats and hamsters. This data set seems to support the notion that airway branching in humans is overall symmetric. 

\textit{Geometry F}. The symmetrical dichotomous branching model~\cite{wiggs1990model} generalizes the  model B~\cite{weibel1963morphology, weibel2014morphometry} by incorporating the mechanisms involved in airway narrowing due to shortening of the smooth muscle. 

\textit{Geometry G}. The asymmetric model~\cite{Olson1970} postulates that the bronchial tree possesses an optimal topology that serves to provide a best physiological function. The chosen function is the lung's ability to move the respiratory gases. The subsequent studies, some of which are referenced in the introduction, suggest a number of alternative lung functions as optimization target. 

\textit{Geometry H}. The asymmetric model~\cite{sturm2012theoretical} provides a statistical representation of the tracheobronchial tree. The model consists of generation-specific probability density functions for airway diameters, airway  lengths, branching angles, and gravity angles. In each generation, the vessel diameter is related to its length $D = 2.115 \ln L + 0.267$.

\subsection{Model of Oxygen and Carbon Dioxide Transport}

\begin{figure}[htbp]
  \centering
  \includegraphics[width=0.9\textwidth]{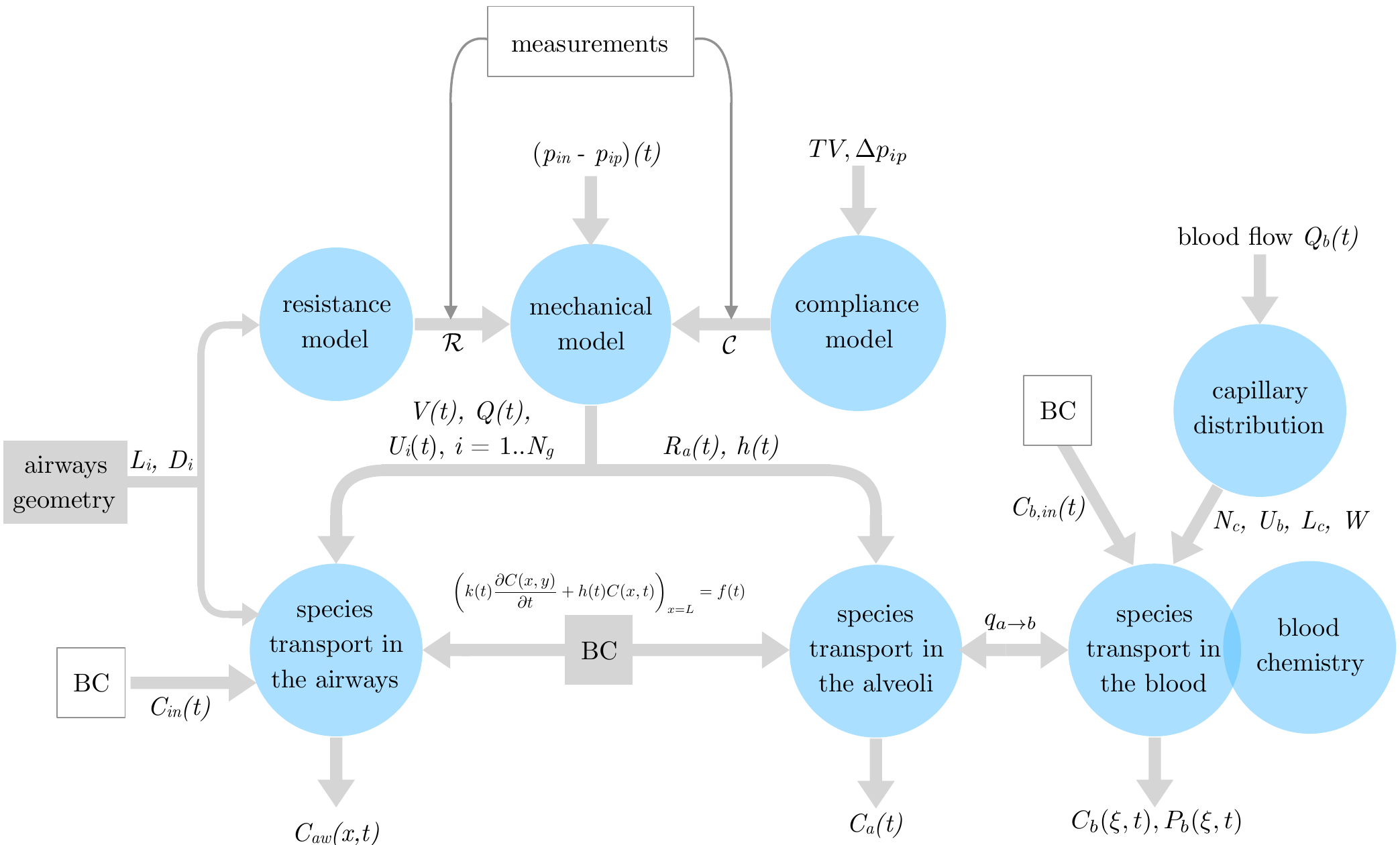}
  \caption{Model workflow showing connections between constituent sub-models, as well as data streams / model inputs.}
  \label{figitALL}
\end{figure}

Respiration is a complex process that starts with ventilation via airways that is governed by lung mechanics, followed by species (mainly oxygen O$_2$ and carbon dioxide CO$_2$) transport across the alveolar-capillary barrier between the alveoli and the circulating blood in the pulmonary capillaries, constrained by the O$_2$-CO$_2$ binding properties of blood. Our model accounts for these phenomena, as shown in Figure \ref{figitALL}. The following assumptions underpin our analysis:
\begin{description}
\item (i) the airways branches are cylinders of circular cross section, 
\item (ii) all the branches belonging to a certain generation are identical, 
\item (iii) all the alveoli are identical, 
\item (iv) the blood capillaries on the circumference of the alveolus are the same for all alveoli, and 
\item (v) the O$_2$ and CO$_2$ concentrations at the capillary inlet are the same for all capillaries. 
 \end{description}
 
\subsubsection{Model of air flow}\label{sec:Mechanical  model}

{
During inspiration/expiration, the lung expands/contracts as air is drawn from/to the atmosphere through the conducting airways to/from the alveolar space~\cite{Guyton}. The expansion  and contraction take place predominantly in the alveoli owing to the negligible contribution of the airways to the lung compliance. Pulmonary ventilation, quantified in terms of the volumetric flow rate $Q(t)$ of the air exchange between the atmosphere and the alveoli, is driven by the time-varying negative intrapleural pressure established by the chest muscles and the recoil of elastic tissues. On the other hand, the resistance to air flow, $\mathcal R$, is dominated by  viscous forces in the airways.
}

\begin{figure}[htbp]
  \centering
  \includegraphics[width=\textwidth]{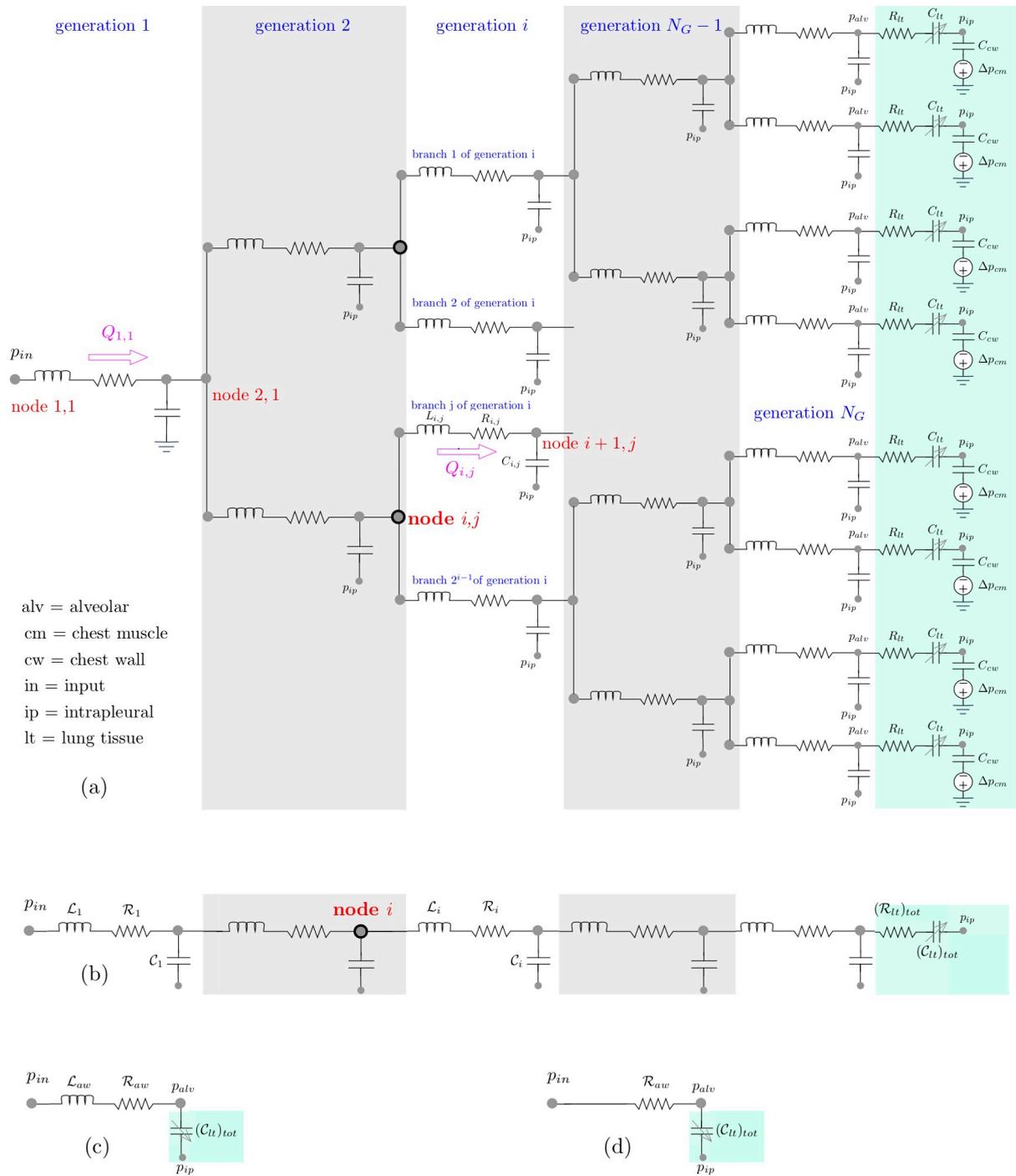}
  \caption{Alternative circuit models of gas flow in the bronchial tree: a general model (a) and its successive simplifications (b--d).}
  \label{figgeneralMechanicalModel}
\end{figure}

We use a circuit representation of flow in airways (Fig.~\ref{figgeneralMechanicalModel}a), such that each branch is conceptualized as a resistor in a series to represent friction; and inertance and shunt capacitance are used to represent the effects of inertia and storage, respectively. The capacitive behavior arises when bronchus wall is compliant.  If all bronchi of the like generation are identical and if the lung tissue resistance and compliance are uniformly distributed over the alveoli, then all the nodes in a branch have the same potential (pressure) as the corresponding nodes in the all the branches belonging to the same generation. In that case, the general airways circuit model in Fig.~\ref{figgeneralMechanicalModel}a simplifies to its counterpart in Fig.~\ref{figgeneralMechanicalModel}b, such that the $(i,j)$ nodes of the general circuit ($j=1,\ldots,2^{i-1}$) are represented by a single $i$ node in the simplified circuit. 

If bronchus wall compliance is small, then bronchi can be approximated as rigid vessels and all the capacitors 
in the network can be replaced by open circuits. This assumption is justified because the lung tissue compliance, associated with expansion and contraction of the alveoli, is much larger than that associated with expansion and contrac\-tion of the bronchi~\cite{mead1969contribution}. 
The tissue resistance is typically much smaller than the airways resistance \cite{ferris1964partitioning} and, hence, can be neglected. 
Under these assumptions, the circuit model in Fig.~\ref{figgeneralMechanicalModel}b simplifies to that in Fig.~\ref{figgeneralMechanicalModel}c. 
Finally, if the inertia effects are negligibly small, then the lung mechanics is represented by the 
circuit in Fig.~\ref{figgeneralMechanicalModel}d.

\begin{figure}[htbp]
\centering
\includegraphics[width=0.45\textwidth]{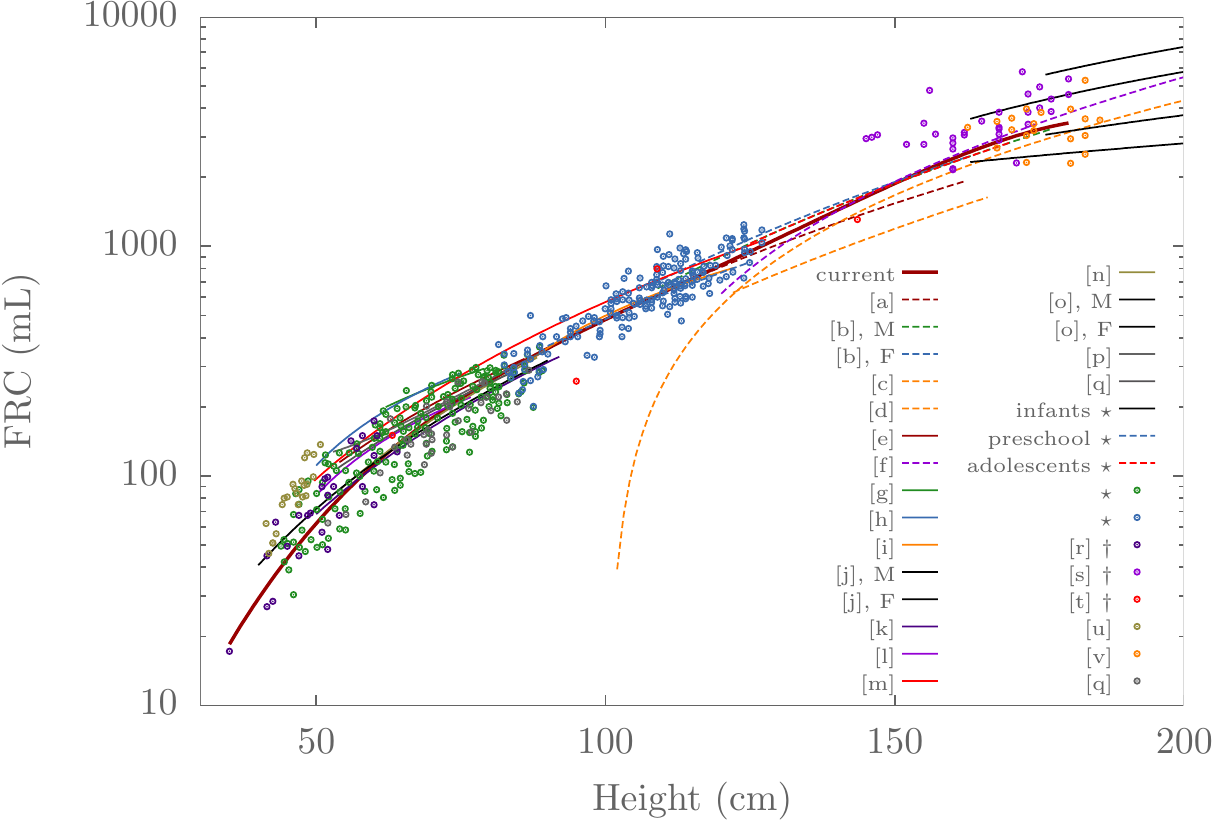}
\includegraphics[width=0.45\textwidth]{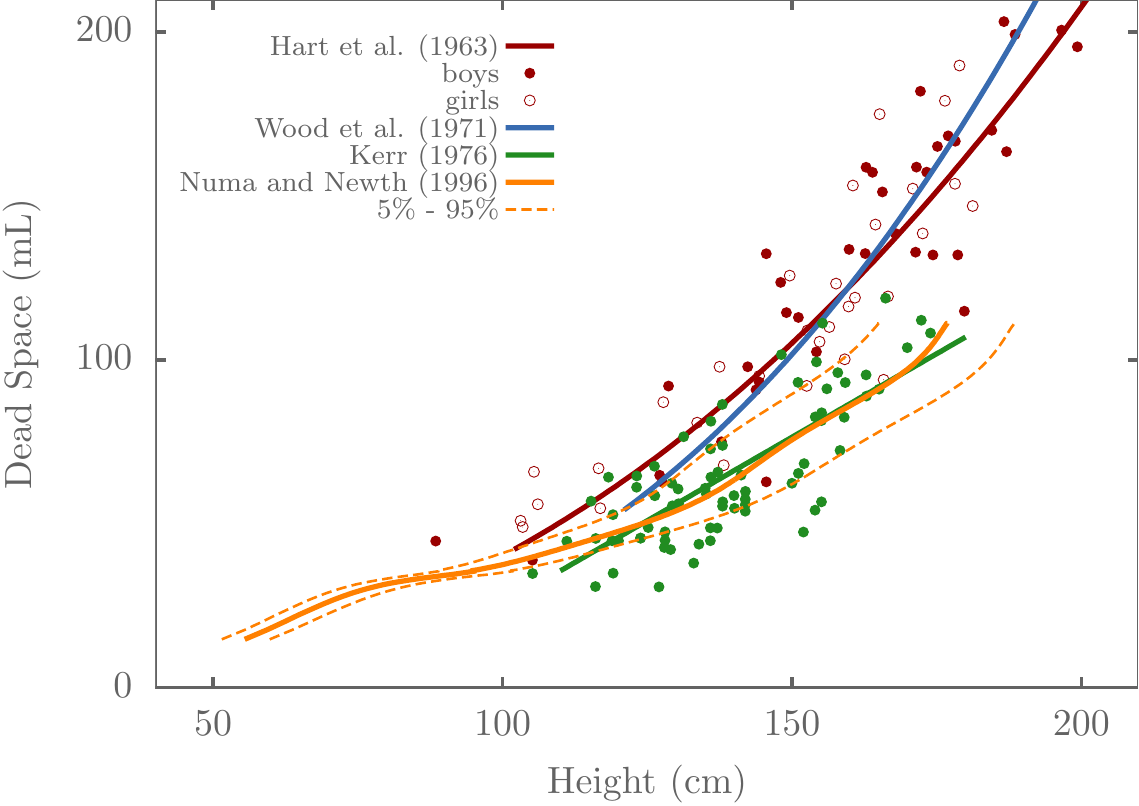}
\caption{Dependence of functional residual capacity FRC (left) and dead space DS (right) on body height, which serves as a proxy for age. The correlation between FRC and lung volume/age reported in~\cite{Thurlbeck1982} were used to convert total lung volume into FRC. When needed, growth charts were used to express height as function of age. [a]: Engstom et al.~(1956), [b]: Cook and Hamann~(1961), [c]:  Engstrom et al.~(1962), [d]: Hart et al.~(1963), [e]: Doershuk et al.~(1970), [f]: Wood et al.~(1971), [g]: Hatch and Taylor~(1976), [h]: Stocks and Godfrey~(1977), [i]:  Greenough et al.~(1986), [j]: Roberts et al.~(1991), [k]: Tepper et al.~(1993), [l]: Merth et al.~(1995), [m]: Stocks and Quanjer~(1995), [n]: McCoy et al.~(1995), [o]: Neder et al.~(1999), [p]: Castile et al.~(2000), [q]: Hulskamp et al.~(2003), [r]:  Hislop et al.~(1986), [s]: Angus and Thurlbeck~(1972), [t]:  Davies and Reid~(1970), [u]:  Auld et al.~(1963), [v]:  Hart et al.~(1963). $\star$: Stocks and Quanjer~(1995) using Tepper et al.~(1993); Hanrahan et al.~(1990); Merth et al.~(1995); Taussig et al.~(1977); Greenough et al.~(1986).}
\label{fig:FRCV_DS}
\end{figure}

\begin{figure}[htbp]
\centering
 \includegraphics[scale=0.9]{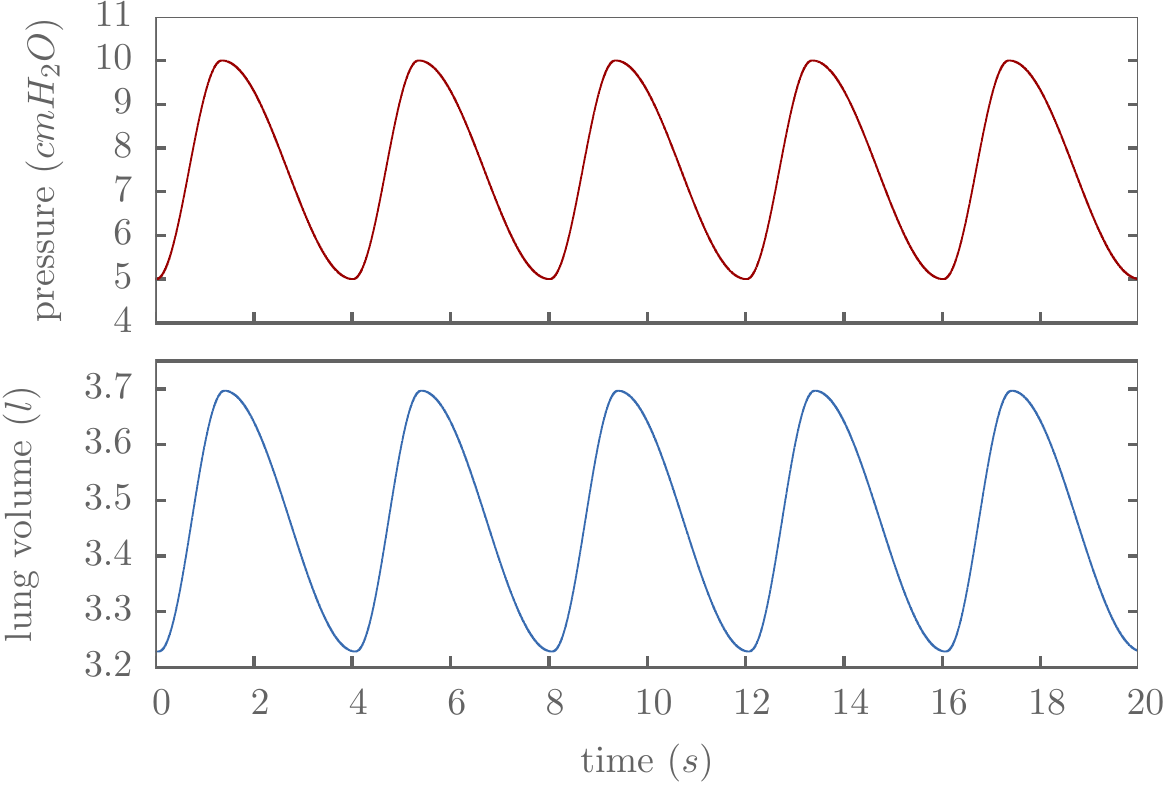}
\caption{Temporal variation of the pressure signal $\Delta p(t)$ {(the model's input)} and the lung volume $V_\text{lu}(t)$ {(the model's output)} under normal breathing conditions for an adult.}
\label{FigDPVNormal}
\end{figure}

For the latter model, the relationship between the (known) pressure difference between the airways inlet and the intrapleural fluid, $\Delta p(t)$, and the (unknown) lung volume, $V_\text{lu}(t)$, is \cite{carvalho2011respiratory} 
\begin{eqnarray}
{\cal R} \frac{\text dV_\text{lu} }{ \text dt} + \frac{V_\text{lu}- V^*}{{\cal C}} = \Delta p,
\label{eqMECHMODEL}
\end{eqnarray} 
where ${\cal C}$ is the compliance of the lung tissue; ${\cal R}$ is the airways resistance; and {$V^* = \text{FRC}  - {\cal C} p_\text{ip,max}$, where the functional residual capacity (FRC) and the maximal intrapleural pressure $p_\text{ip,max}$ are model parameters. For normal breathing of an adult, $\mathcal R = 0.56$~cmH$_2$O s/l and ${\cal C} = 93.66$~ml/cmH$_2$O~\cite{Ochs2004}, and $p_\text{ip,max} = - 5$~cmH$_2$O~\cite{Guyton}. Values of the FRC, collected from multiple prior studies, are shown in Fig.~\ref{fig:FRCV_DS} (left) as function of body height which serves as a proxy for age. The pulsating pressure signal $\Delta p(t)$ is generated via a relation
\[ \Delta p = -p_\text{ip,max} + \frac{p_\text{ip,max}-p_\text{ip,min}}{2}
\begin{cases}
     1 + \sin \left( -\dfrac{\pi}{2} + \dfrac{\pi f t'}{i_f} \right)       & \quad \text{for } 0 \leq t' \leq i_f T \\ \\
     1 + \sin \left( \dfrac{\pi}{2} + \dfrac{\pi f t'}{e_f} \right)     & \quad \text{for } i_f T < t' < T
  \end{cases}
\]
where $i_f = (I/E) / (1+I/E)$, $e_f = 1 - i_f$, $t'=t - T \lfloor t/T \rfloor$ and $T=1/f$, with $\lfloor \cdot \rfloor$ denoting the floor function.
In the simulations reported below, we set $p_\text{ip,min} = - 10$~cmH$_2$O~\cite{Guyton}, the adult normal breathing frequency to $f = 1/T = 0.25$~Hz (15 bpm), and the inspiration-to-expiration ratio to $I/E=$0.5~\cite{strauss2000relative}. At the beginning of inspiration ($t=0$), $V_\text{lu}(0) = \text{FRC}$ and $\Delta p(0) = - p_\text{ip,max}$. With these inputs, a solution to the ordinary differential equation~\eqref{eqMECHMODEL} provides the lung volume response, $V_\text{lu}(t)$, to the pressure signal $\Delta p(t)$ (Fig.~\ref{FigDPVNormal}). It also enables us to compute $Q(t)$ $= {\text dV_\text{lu} }/{ \text dt}$, the total volumetric flow rate of air through the airways.}


%
%

\begin{figure}[htbp]
\centering
\includegraphics[scale=0.65]{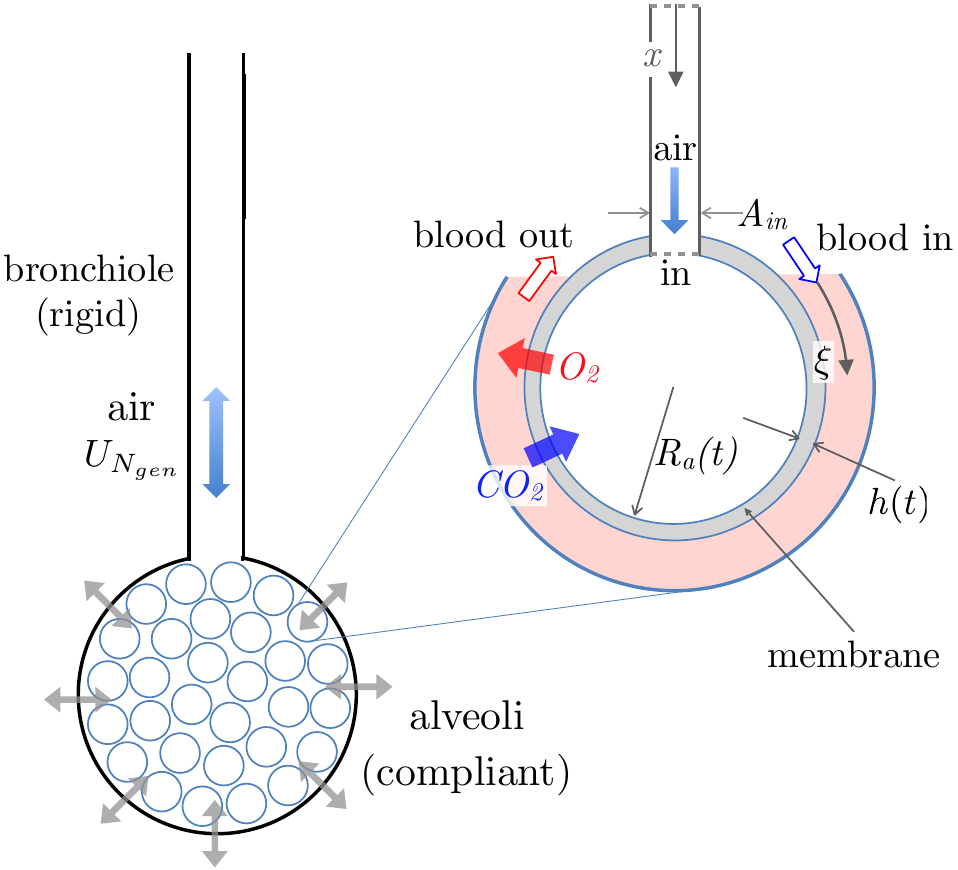}
\caption{Each terminal branch is connected to a number of alveoli. Our model assumes that the bronchiole is connected to the center of each compliant spherical alveolus.}
\label{FIGalveolusSpecies}
\end{figure}

{Next, we partition the total volumetric flow rate $Q(t)$ among individual bronchus of the bronchial tree. The latter is assumed to branch dichotomously and consist of $N_\text{gen}$ generations. That implies that the $i$th generation ($i = 1, \ldots, N_\text{gen}$) comprises $N_i = 2^{i-1}$ bronchi. Each bronchus of the $i$th generation is modeled as a cylinder of diameter $D_i$ and cross-sectional area $A_i = \pi D_i^2/4$. When applied to such a dichotomously branching tree, mass conservation requires the cross-sectionally averaged flow velocity in the $i$th-generation bronchus, $U_i(t)$, to be
\begin{equation}
    U_i(t) = \frac{Q(t)}{N_i A_i}, \qquad i = 1, \ldots, N_\text{gen}.
\end{equation}
This expression implies that the average velocity in the last generation, $U_{N_\text{gen}} = Q / (N_{N_\text{gen}} A_{N_\text{gen}})$, is  small but non-zero. The terminal velocity $U_{N_\text{gen}}$ is the velocity of air through the terminal bronchioles as air fills/empties the expanding/contracting alveoli (Fig.~\ref{FIGalveolusSpecies}).}

\begin{figure}[htbp]
\centering
\includegraphics[width=0.44\textwidth]{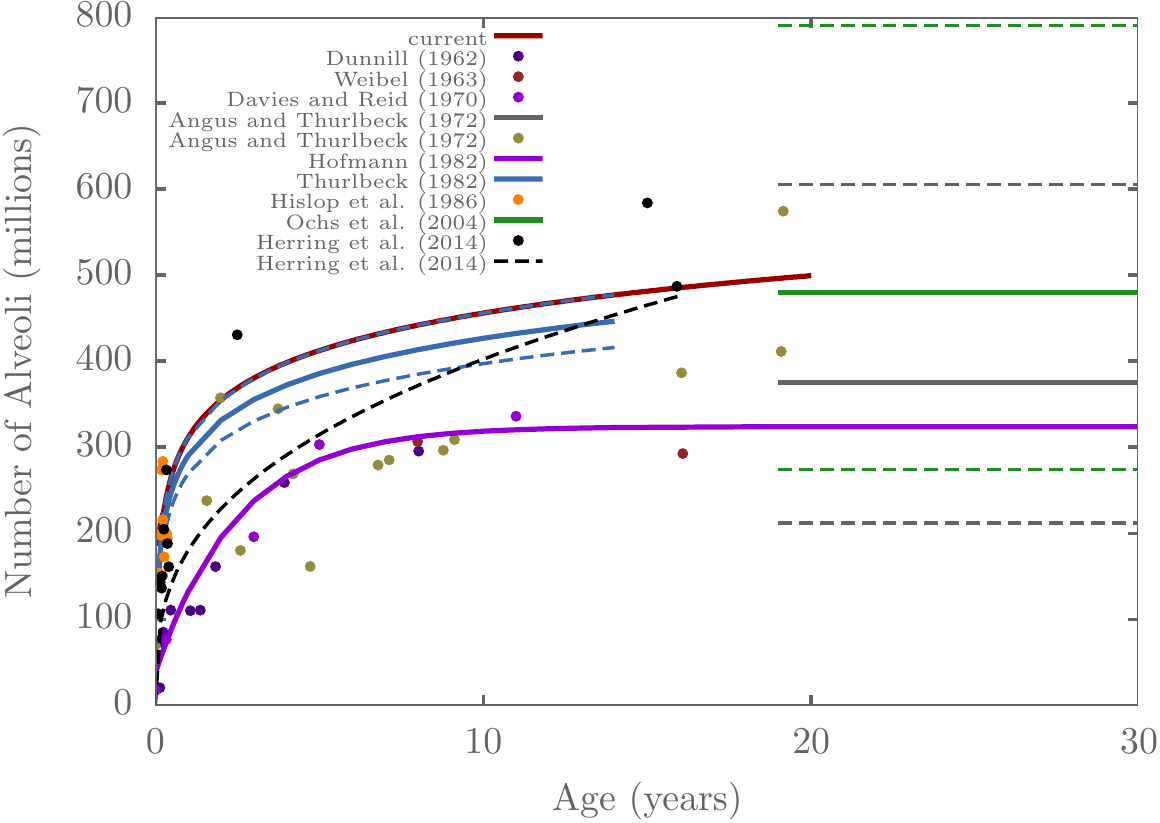}
\includegraphics[width=0.44\textwidth]{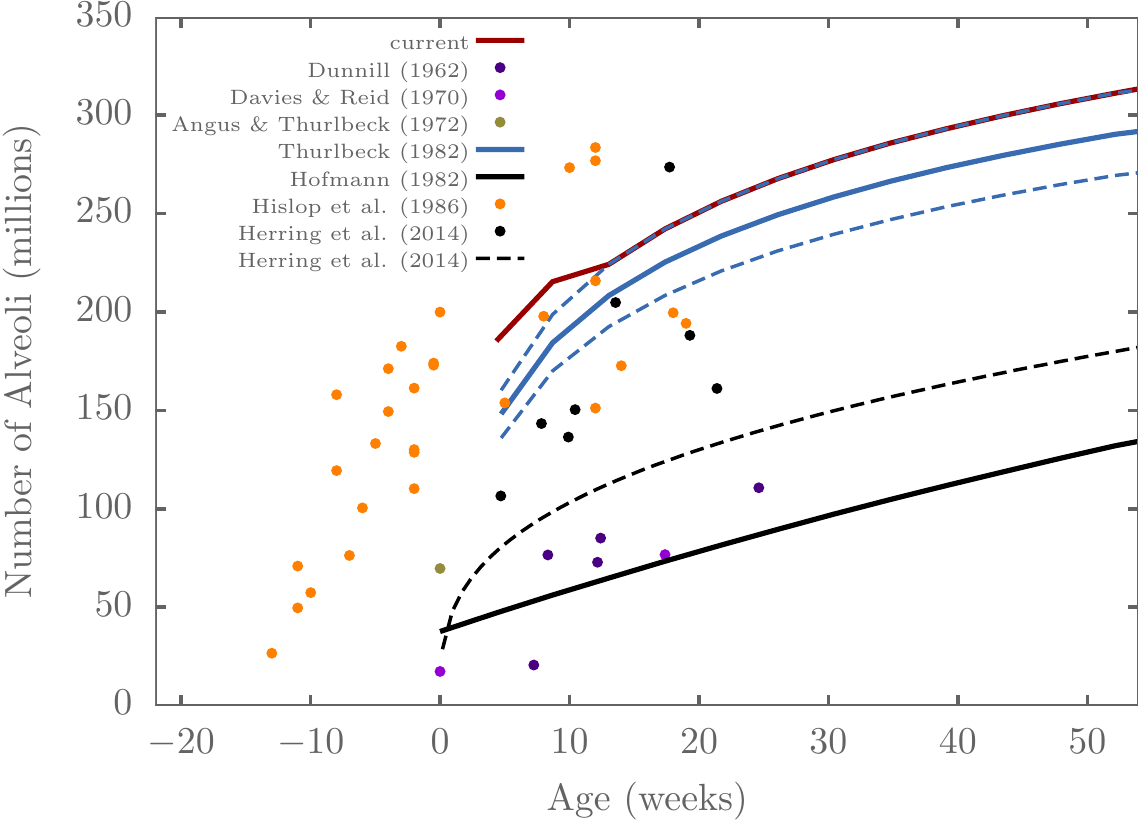}
\caption{Dependence of the number of alveoli, $N_\text{al}$, on age of adults (left) and infants (right).}
\label{fig:Nal}
\end{figure}

Each bronchiole (terminal airways branch) is connected to $N_\text{al}^\text{tb} = N_\text{al} / 2^{N_\text{gen}-1}$ alveoli (Fig.~\ref{FIGalveolusSpecies}). The total number of alveoli depends on age, sex, and lung volume, ranging from $N_\text{al} \approx 3 \cdot 10^7$ in a preterm infant to $N_\text{al} \approx 5 \cdot 10^8$ in a twenty year old adult~\cite{Thurlbeck1982, Ochs2004, herring2014growth} (Fig.~\ref{fig:Nal} and Table \ref{tableModelParameters2}).  For an adult, assuming $N_\text{gen} = 24$, this yields  the number of alveoli per bronchiole to be approximately 60. We treat alveoli as compliant spheres of radius $R_\text{al}(t)$; their total volume is $N_\text{al} V_\text{al}(t)$, where the volume of a single alveolus is $V_\text{al} = 4 \pi R_\text{al}^3 / 3$. {At the beginning of inspiration ($t=0$), $R_\text{al}(0) = 100$~$\mu$m~\cite{Ochs2004}}. 
Since the lung volume $V_\text{lu}(t) = N_\text{al} V_\text{al}(t) + \text{DS}$ is the sum of the total alveolar volume, $N_\text{al} V_\text{al}(t)$, and the time-invariant dead space volume, $\text{DS}$, the change of  $V_\text{al}$ with time satisfies 
\begin{equation}\label{eq:Val}
    \frac{\text dV_\text{al}}{\text dt} = \frac{Q(t)}{N_\text{al}}.
\end{equation}
At the end of expiration, the total alveolar volume is 
$N_\text{al} V_\text{al}(t=0) = \text{FRC} - \text{DS}$.
Values of DS, collected from multiple prior studies, are shown in Fig.~\ref{fig:FRCV_DS} (right) as function of body height, which serves as a proxy for age. {With this initial condition, a solution of~\eqref{eq:Val} yields $V_\text{al}(t)$. }

\subsection{Model of transport of dissolved species }


\subsubsection{Gas transport in airways}

%
%
At any time $t > 0$ and location $x \in (0,L_i)$ along the $i$th bronchus, cross-sectionally averaged volumetric concentration of gases (O$_2$ or CO$_2$), $C_\text{aw}^{(i)}(x,t)$, satisfies an advection-dispersion equation~\cite{zimmerman-2018-hydrodynamic}
\begin{align}\label{EQADVDIFF}
\frac{\partial C_\text{aw}^{(i)}}{\partial t} + U_i(t) \frac{\partial C_\text{aw}^{(i)}}{\partial x} = \mathcal{D}_i \frac{\partial^2 C_\text{aw}^{(i)}}{\partial x^2}, \qquad i=1,\cdots,N_\text{gen},
\end{align}
{where $\mathcal D_i = \mathcal D_\text{aw} + D_i^2 U_i^2/(192 \mathcal D_\text{aw})$ is the hydrodynamic dispersion coefficient, and $\mathcal D_\text{aw}$ is the coefficient of molecular diffusion of the corresponding species (O$_2$ or CO$_2$). The equations for adjacent bronchi are coupled to ensure mass conservation. In the inspiration phase, this coupling gives rise to boundary conditions
\begin{equation}\label{EQADVDIFFBCS}
\begin{split}
& C_\text{aw}^{(1)}(0,t) = C_\text{aw}^\text{in}; \quad 
\text{$C_\text{aw}^{(i+1)}(0,t) = C_\text{aw}^{(i)}(L_i,t)$,  $i=1,\ldots,N_\text{gen}-1$}; \quad
C_\text{aw}^{(N_\text{gen})}(L_{N_\text{gen}},t) = C_\text{al}^\text{in} \\
& \mathcal D_i A_i \frac{\partial C_\text{aw}^{(i)}}{\partial x}(L_i,t)  = 2 \mathcal D_{i+1} A_{i+1} \frac{\partial C_\text{aw}^{(i+1)}}{\partial x}(0,t), \;  i = 1,\ldots,N_\text{gen}-1;
\end{split}
\end{equation}
where $C_\text{aw}^\text{in}$ is the (prescribed) concentration of O$_2$ or CO$_2$ at the inlet to trachea ($i=1$), and $C_\text{al}^\text{in}$ is the concentration of O$_2$ or CO$_2$ at the interfaces between the terminal bronchi ($i = N_\text{gen}$) and the alveoli. Since the latter is unknown, Eq.~\eqref{EQADVDIFF} with $i= N_\text{gen}$ is supplemented with a Robin boundary condition at $x = L_{N_\text{gen}}$, Eq.~\eqref{EQSPECIESALV1}, which accounts for the gases exchange between the alveolar space, the terminal bronchioles, and the blood (see below).
}

{In the expiration phase, the first of the boundary conditions~\eqref{EQADVDIFFBCS} is replaced with $\partial_x C_\text{aw}^{(1)}(0,t) = 0$.
}

\subsubsection {Gas transport in alveoli}

The alveoli adjacent to a bronchiole are interconnected via ``pores of Kohn''~\cite{desplechain1983pores}, and form a sac (Fig.~\ref{FIGalveolusSpecies}). {These interconnections justify modeling gases in all the alveoli belonging to a single sac as well mixed, so that $C_\text{al} = C_\text{al}^\text{in}$ for any single alveolus. 
This assumption is appropriate because the time scale associated with diffusive transport over the alveolar radius ($\sim 0.7$~ms for adults, and $\sim 0.07$~ms for infants) is much smaller than the respiration time scale. 
} Mass conservation of dissolved gases in the alveolar space  yields a mass-balance equation for the average concentration of the species in the alveolar space, $C_\text{al}(t)$,
\begin{equation}
\frac{\text d C_\text{al}V_\text{al} }{ \text dt}={Q}_\text{al}^\text{in} C_\text{al}^\text{in} + A_\text{al}^\text{in} J_\text{diff} - A_\text{ca} q_{\text{al} \rightarrow \text{bl}},
\label{EQSPECIESALV}
\end{equation}
where $Q_\text{al}^\text{in} = Q(t)/N_\text{al}$; 
{$A_\text{al}^\text{in} = A_{N_\text{gen}}/N_\text{al}^\text{tb}$ is the cross-sectional area of the terminal bronchiole per connected alveolus};  $A_\text{ca}$ is the capillary-alveolus contact area; $q_{\text{al} \rightarrow \text{bl} }$ is the mean species flux from the alveolar space into the blood; and {$J_\text{diff} = - \mathcal D_{N_\text{gen}} \partial_x C_\text{aw}(L_{N_\text{gen}},t)$} is the diffusive flux of the dissolved species from the terminal bronchiole into the alveolus. {Since $Q_\text{al}^\text{in} N_\text{al}^\text{tb} = Q/N_\text{gen}$, equation~\eqref{EQSPECIESALV} yields the Robin boundary condition at the end of the terminal bronchioles
\begin{equation}
\mathcal D_{N_\text{gen}} A_{N_\text{gen}} \frac{\partial C_\text{aw}^{(N_\text{gen})}}{\partial x}(L_{N_\text{gen}},t) + \frac{Q}{N_\text{gen}} C_\text{aw}^{(N_\text{gen})}(L_{N_\text{gen}},t) = N_\text{al}^\text{tb} \left( \frac{\text d C_\text{al}V_\text{al} }{ \text dt} + A_\text{ca} q_{\text{al} \rightarrow \text{bl}} \right).
\label{EQSPECIESALV1}
\end{equation}
This completes the specification of the auxiliary conditions for~\eqref{EQADVDIFF}, which couple transport of O$_2$ and CO$_2$ to the physiological processes in pulmonary membrane and blood through (as yet unspecified) quantities $C_\text{al}$ and $q_{\text{al} \rightarrow \text{bl}}$.
}

\subsubsection {Gas transport in pulmonary membrane}

Exchange of gases between the alveolar air and the blood circulating in the surrounding capillaries occurs by diffusion through a pulmonary membrane, whose thickness is $h_0 \approx 0.6$~$\mu$m~\cite{weibel08}. Diffusion across the membrane is driven by a difference in partial pressure between the blood side and the alveolar side. Partial pressure of O$_2$ is typically higher in the alveolus, while partial pressure of CO$_2$ is higher on the blood side. Hence, O$_2$ diffuses from the alveoli into the capillaries, and CO$_2$ diffuses from the capillaries into the alveoli (see Fig.~\ref{FIGalveolusSpecies}). Since the alveolus radius is two orders of magnitude larger than the membrane thickness, the membrane is assumed to be locally planar, so that the species concentration in the membrane, $C_\text{mem}(r,t)$, is described by a diffusion equation
$\partial_t C_\text{mem} = {\cal D}_\text{wat} \partial_r^2 C_\text{mem}$,
where ${\cal D}_\text{wat}$ is the diffusion coefficient in the membrane, which is set to that in water. Since the diffusion time scale, $h_0^2/{\cal D}_\text{wat} \sim 1$~ms for both O$_2$ and CO$_2$, is much smaller than the respiration period (1.5 -- 4~s), we assume diffusion through the membrane to be quasi-steady so that the species flux through the membrane is uniform
\begin{equation}
q_{\text{al} \rightarrow \text{bl}}(\xi, t)=\frac{{\cal D}_\text{wat}}{k_\text{H}} \frac{{C}_\text{al}(t)- P_\text{bl}(\xi, t) / (RT)}{h(t)}.
\label{EqMembraneDiff}
\end{equation}
Here $P_\text{bl}(r, t)$ is the species (O$_2$ or CO$_2$) partial pressure in the blood at point $\xi$ along the capillary, $k_\text{H}$ is Henry's constant, $R$ is universal ideal gas constant, and $T$ is the temperature. 
The alveolar membrane of thickness $h(t)$ is viscoelastic and responds to any change in alveolar pressure in a way that conserves the volume of the membrane, $V_\text{mem}$:
\begin{equation}
h(t)=\sqrt[3]{\frac{3}{4\pi}\left[ {V}_\text{mem}+\frac {4}{3}\pi{R}_\text{al}^3(t)\right]}-{R}_\text{al}(t).
\end{equation}

On the blood side, the partial pressure $P$ and concentration $C$ are related by the dissociation curves, presented in the next section.

\subsubsection {Gas transport in blood}
Each alveolus is surrounded by several pulmonary capillaries, in which the red blood cells travel at the blood speed ${U}_\text{bl}$. The diameter of the capillary, $D_\text{cap} =  10$~$\mu$m, equals the size of a red blood cell. Nevertheless, as in \cite{liu98}, our blood perfusion model treats blood in the capillaries as a uniform homogeneous phase (masking the discrete constituents: plasma and erythrocytes). The model assumes equilibrium conditions for the reactions, thus allowing the use of an empirical dissociation curves to relate the species' partial pressure to its concentration. If the speed of blood along the capillary is $U_\text{bl}$, then the concentrations of both O$_2$ and CO$_2$ in blood, at any space-time point $(\xi,t)$ along a capillary, are described by
{
\begin{equation}
\frac{\partial C_{\text{bl},k}}{\partial t} + U_\text{bl} \frac{\partial C_{\text{bl},k} }{\partial \xi}=\frac{W}{{A}_\text{cap}}{q}_{\text{al},k \rightarrow \text{bl} }, \qquad k = \text{O$_2$ and CO$_2$}.
\label{eqO2Blood}
\end{equation}}
Here $A_\text{cap} = \pi {D}_\text{cap}^2 / 4$ is the cross-sectional area of the capillary, and $W$ is the contact area (per unit length of the capillary) between the capillary and the alveolar membrane.  Gas exchange between the alveolar space and the blood circulating in the surrounding capillaries is a function of the blood speed in the capillaries, the contact area between an alveolus and the surrounding capillaries, and the capillaries distribution. 

The blood speed in a single pulmonary capillary, $U_\text{bl}$, is estimated as
\begin{equation}\label{eq:Ubl}
U_\text{bl}=\frac{ \dot{V}_{\text{bl} / \text{al} } }{ N_{\text{cap}/\text{al}}{A}_\text{cap}}.
\end{equation}
The number of capillaries per alveolus is
\begin{equation*}
N_{\text{cap}/\text{al}} = \frac{1 }{ A_\text{cap} L_\text{cap} } \left(\frac{V_\text{bl} }{ N_\text{al} } \right),
\end{equation*}
where, assuming that a single capillary extends over half of the alveolar perimeter, $L_\text{cap} = \pi R_{\text{al},0}$; and, assuming that each alveolus is surrounded by a spherical shell of thickness equal to $D_\text{cap}$, the blood volume in the pulmonary capillaries is $V_\text{bl} = \beta V_\text{bl, max}$, with $V_\text{bl, max} = N_\text{al} 4\pi{R}_{\text{al},0}^2 D_\text{cap}$ providing the upper bound on the blood volume in the capillaries, and $\beta$ denoting the percent contact area between the alveoli and the capillaries. Finally, in~\ref{eq:Ubl}, the capillaries blood flow rate per alveolus is $\dot{V}_{\text{bl} / \text{al} } = \dot{V}_\text{bl} / N_\text{al}$, where $\dot{V}_\text{bl} = V_\text{bl} / \tau_\text{r}$ is the volume flow rate of blood in the pulmonary capillaries and $\tau_\text{r}$ is the residence time of blood in a pulmonary capillary, with a typical value of $\tau_\text{r} \approx 1$~s  \cite{warren1991red}.

Equations~\ref{eqO2Blood} are coupled through the interdependence between O$_2$ and CO$_2$, which arrises due to the blood chemistry.
We use the O$_2$-hemoglobin dissociation curve~\cite{Sharan1989}, derived from a kinetic study of  several chemical reactions between the respiratory gases and hemoglobin, to relate the O$_2$ concentration, $C_{\text{O}_2}$, to the O$_2$ partial pressure, $P_{\text{O}_2}$; the reaction rate constant in this relation depends on pH and the CO$_2$ partial pressure.
%
We also use the CO$_2$ dissociation curve~\cite{Meade1972} to relate the CO$_2$ concentration, $C_{\text{CO}_2}$ to both its partial pressure, $P_{\text{CO}_2}$, and the O$_2$ partial pressure, $P_{\text{O}_2}$.

These two dissociation curves capture both Bohr's and Haldane's effects in blood chemistry, both of which stem from the coupling between O$_2$ and CO$_2$. Bohr's effect refers to the observation that, for a fixed value of $P_{\text{O}_2}$, the hemoglobin saturation and, hence, $C_{\text{CO}_2}$ increase as $P_{\text{CO}_2}$ decreases. Haldane's effect refers to the observation that, for a fixed value of $P_{\text{CO}_2}$, both the hemoglobin saturation and $C_{\text{O}_2}$ decrease as $C_{\text{CO}_2}$ increases. 

The boundary conditions for Eqs.~\eqref{eqO2Blood} at the inlet to the capillaries ($\xi = 0$) are set to $P_{\text{O}_2}(\xi = 0,t) = 40$ mmHg and $P_{\text{CO}_2}(\xi = 0,t) = 46$ mmHg.




\section{Numerical Implementation}

\begin{figure}[htbp]
  \centering
  \includegraphics[width=0.9\textwidth]{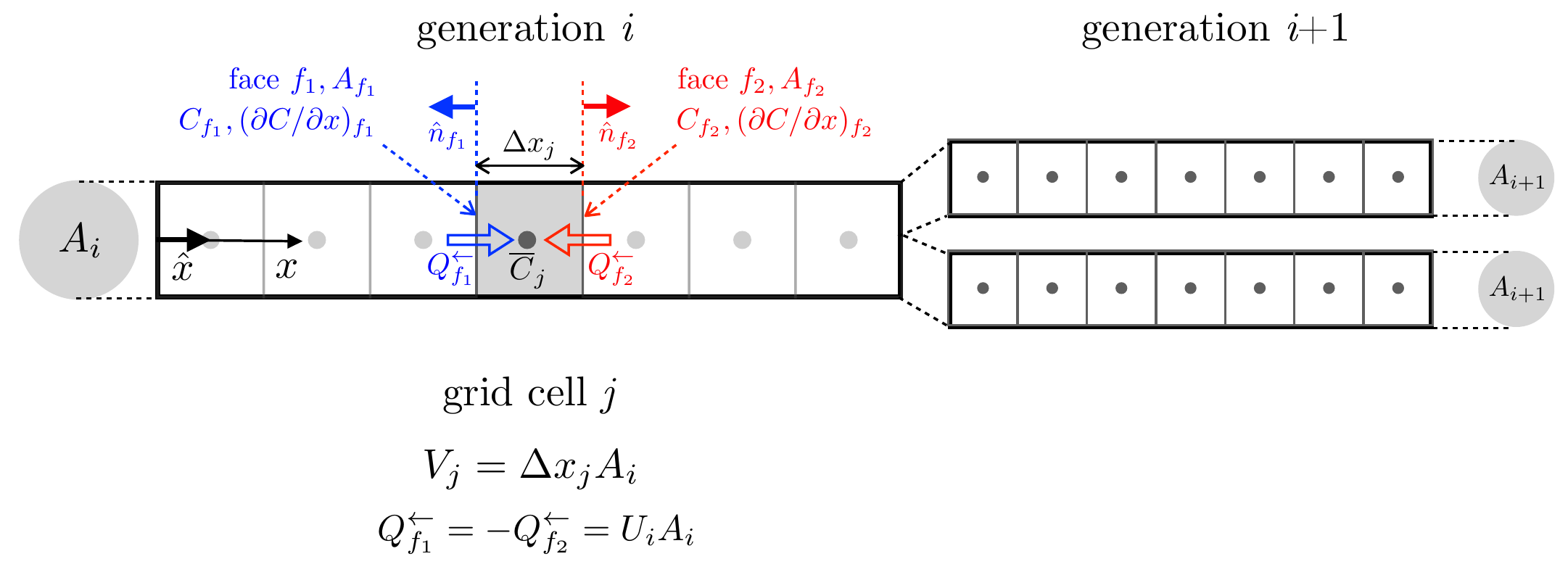}
  \caption{A schematic of the mesh used in the finite volume method for a branch in generation $i$ and the two daughter branches in generation $i+1$.}
  \label{figmesh}
\end{figure}

{
Equations~\eqref{EQADVDIFF} and~\eqref{EQADVDIFFBCS} are solved using the finite volume method~\cite{moukalled2016finite}, which conserves mass by matching the gases' fluxes at the faces of the grid cells. For a fixed control volume $\Omega_j$, e.g., the $j$th grid cell in Fig.~\ref{figmesh}, the advection-dispersion equation~\eqref{EQADVDIFF} is expressed as 
\begin{eqnarray}
V_j \frac{\partial \bar C_j}{\partial t} = \sum_{f \subset \partial \Omega_j } A_f \left( Q_f^{\leftarrow} C_f + \mathcal{D}_f  \frac{\partial C_f}{\partial n_f} \right),
\label{EQADVDIFFCV}
\end{eqnarray}
where $\bar C_j$ is the average concentration in the control volume $\Omega_j$, whose size is $V_j = A_j \Delta x_j$ and whose surface is denoted by $\partial \Omega_j$; the summation is over all the faces $f$ comprising the control surface $\partial \Omega_j$; $Q^{\leftarrow}_f$ is the volumetric flow rate into the control volume across the face $f$; $A_f$ is the area of the face $f$; $C_f$ is the concentration at the face $f$; and the derivative in the diffusion flux is in the direction of the normal vector to the face $f$. 
}

{We use a second-order time integration scheme. The advection and diffusion terms are discretized using a second-order upwind and a second-order central difference schemes, respectively. Since O$_2$ and CO$_2$ are coupled through the oxyhemoglobin dissociation, the numerical solution over each time step is obtained by iterating between the airways and the blood until the concentrations and the fluxes match at the interface. Although not shown here, the grid size and the time step are chosen based on a standard convergence study.}

\section{Results}
 
The input data needed for our model are summarized in Table~\ref{tableModelParameters2} for a human newborn and a twenty-year-old healthy male. Their height and mass are at the 50th percentile line of their respective growth charts. Additional input for our model is the alternative bronchial tree representations reported in Tables~\ref{Table:models2} and~\ref{Table:models3}. In the simulations reported below we compare the model predictions based on these tree representations to ascertain their compatibility and ability to predict key observable physiological quantities. (We excluded the tree geometries~\cite{ionescu2010viscoelasticity} and \cite{horsfield1968morphology} from our simulations because of their unphysically high dead space volumes and unphysically small resistance, as shown in Table~\ref{Table:assessmentTable}).

\begin{table}[htbp]
\begin{center}
\caption{Values of model parameters for two baseline cases.}
\label{tableModelParameters2}
\begin{tabular}{ l  c  c  }
\toprule
Parameter & infant & adult \\
\midrule
Age (yrs) & 0 & 20\\
Gender& male & male \\
Height  (cm) & 49.99&  176.85   \\
Mass (kg) & 3.53 & 70.6 \\
Respiration frequency (Hz) & 0.75& 0.25  \\
Tidal volume (ml) & 21.7 & 468.3\\
Functional residual capacity (ml)  & 61.9 & 3228.27 \\
Initial alveolar radius ($\mu$m) & 35 & 100 \\
Number of alveoli (million) & 152 & 499 \\
Minimum intrapleural pressure (cmH$_2$O) & -10 & -10 \\
Maximum intrapleural pressure (cmH$_2$O) & -5 & -5 \\
Compliance (ml/cmH$_2$O) & 4.34 & 93.66\\
Blood flow rate in capillaries  (l/min) & 0.518  & 11.2 \\
\bottomrule
\end{tabular}
\end{center}
\end{table}

\begin{figure}[htbp]
\centering
  \includegraphics[width=0.49\textwidth]{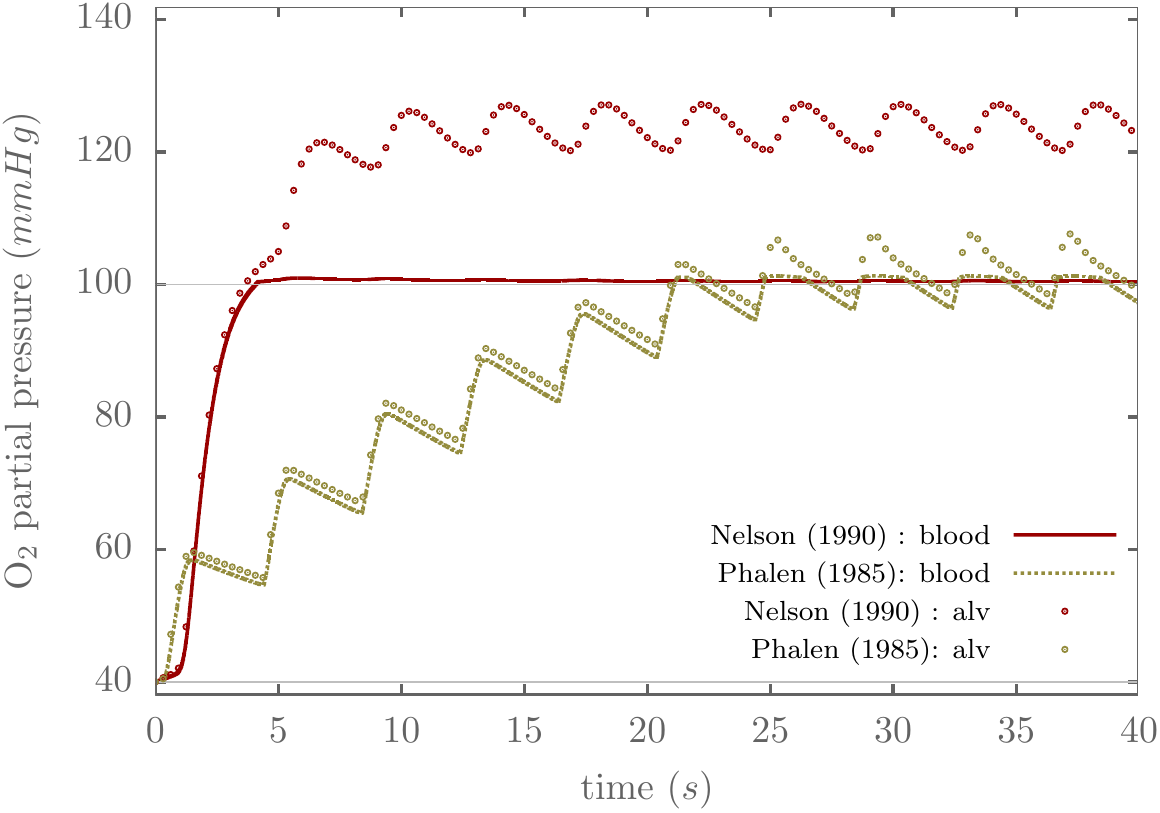}
  \includegraphics[width=0.49\textwidth]{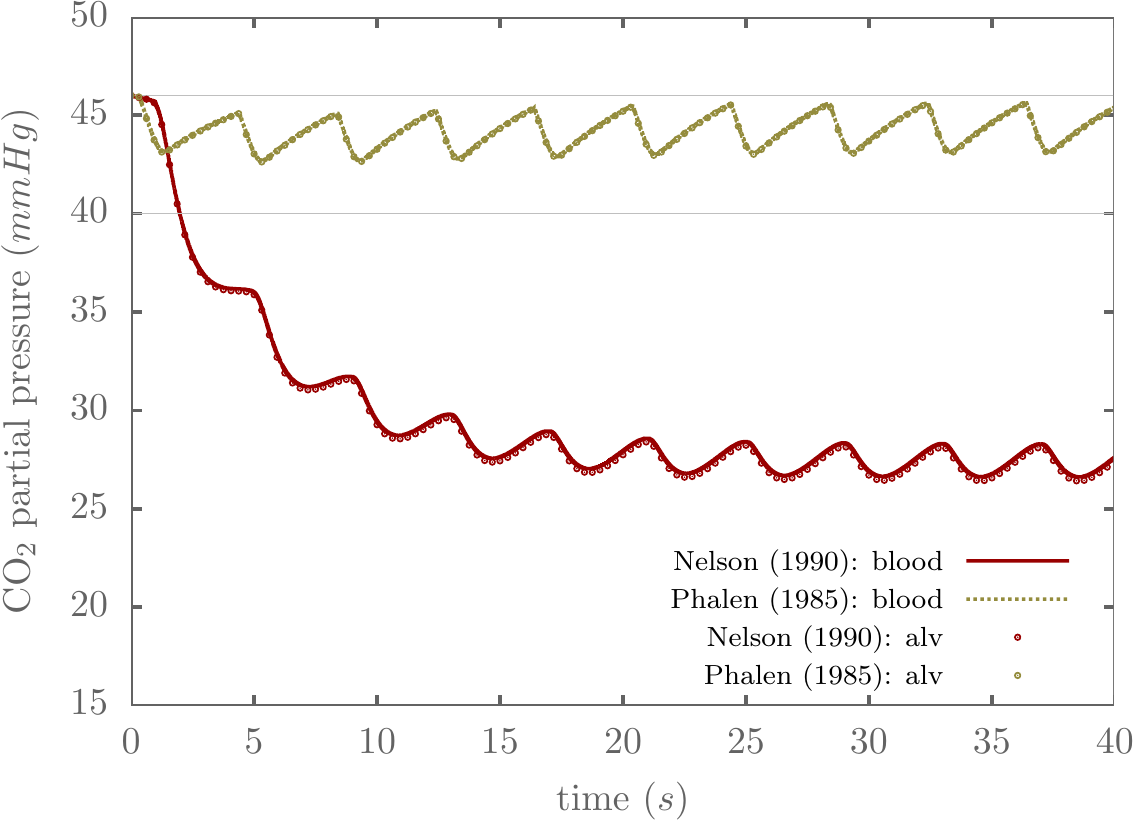}
\caption{Temporal variation of the mean partial pressure of oxygen (left) and carbon dioxide (right) in the blood and in the alveolar space of adults, as predicted with our model for two bronchial tree representations.  }
\label{FigO2CO2BAAdults}
\end{figure}

Since the time scale of diffusion across the alveolar membrane is proportional to the the square of the membrane thickness, it is much smaller than the time scales of the other transport mechanisms. Consequently, gas exchange between the alveolar space and the blood occurs very quickly, so that its partial pressures on both sides remain close to each other, as long as O$_2$ in the blood is below saturation level. Figure~\ref{FigO2CO2BAAdults} shows that this is indeed the case: when the blood is not saturated with oxygen, $P_{\text{O}_2}$ in the alveolar space is close to but smaller than $P_{\text{O}_2}$ in the blood, indicating O$_2$ transport into the blood. This figure also demonstrates that $P_{\text{CO}_2}$ in the blood is close to but larger than $P_{\text{CO}_2}$ in the alveolar space, indicating CO$_2$ transfer into the alveolar space. 

\begin{figure}[htbp]
\centering
  \includegraphics[width=0.49\textwidth]{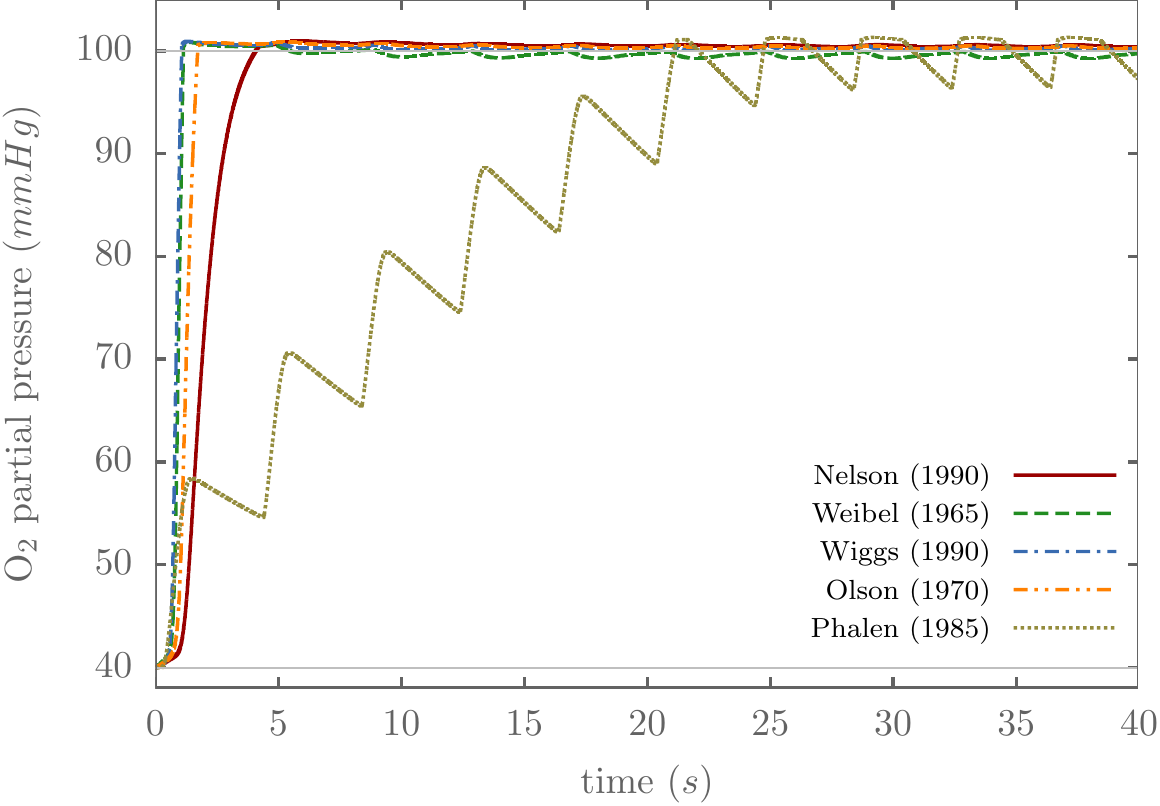}
  \includegraphics[width=0.49\textwidth]{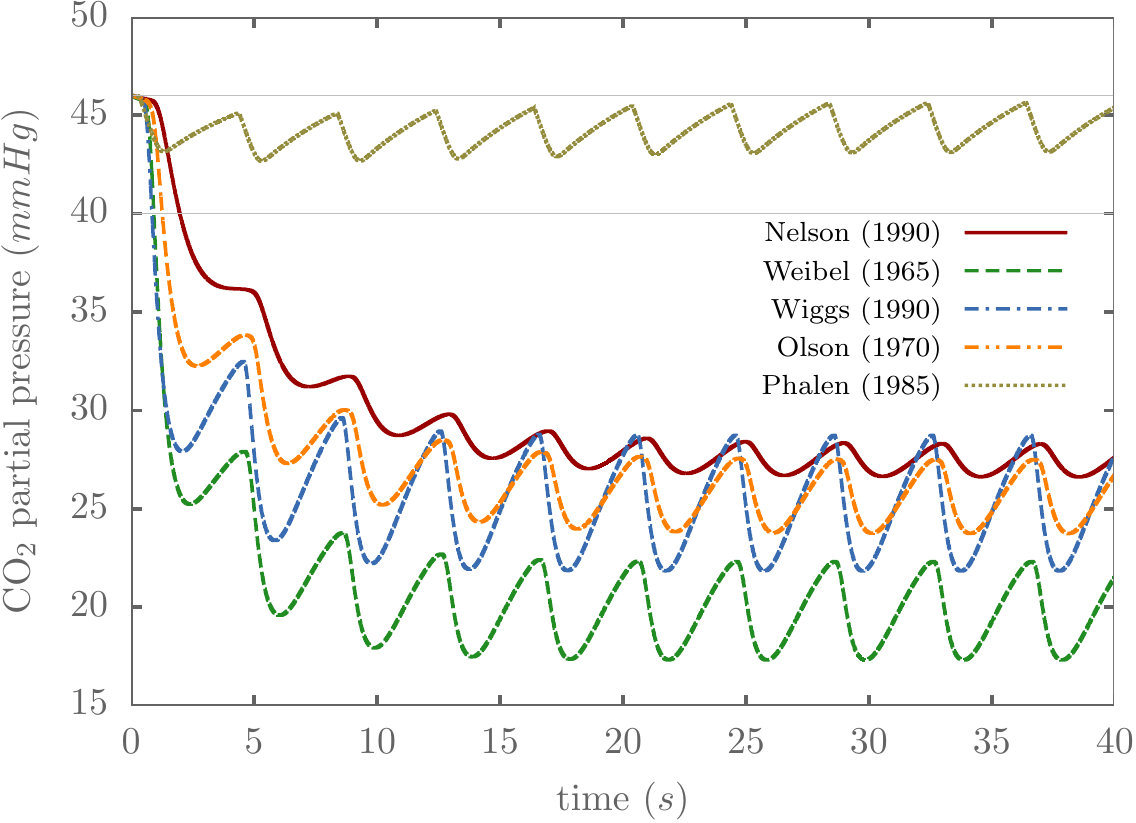}
  \includegraphics[width=0.49\textwidth]{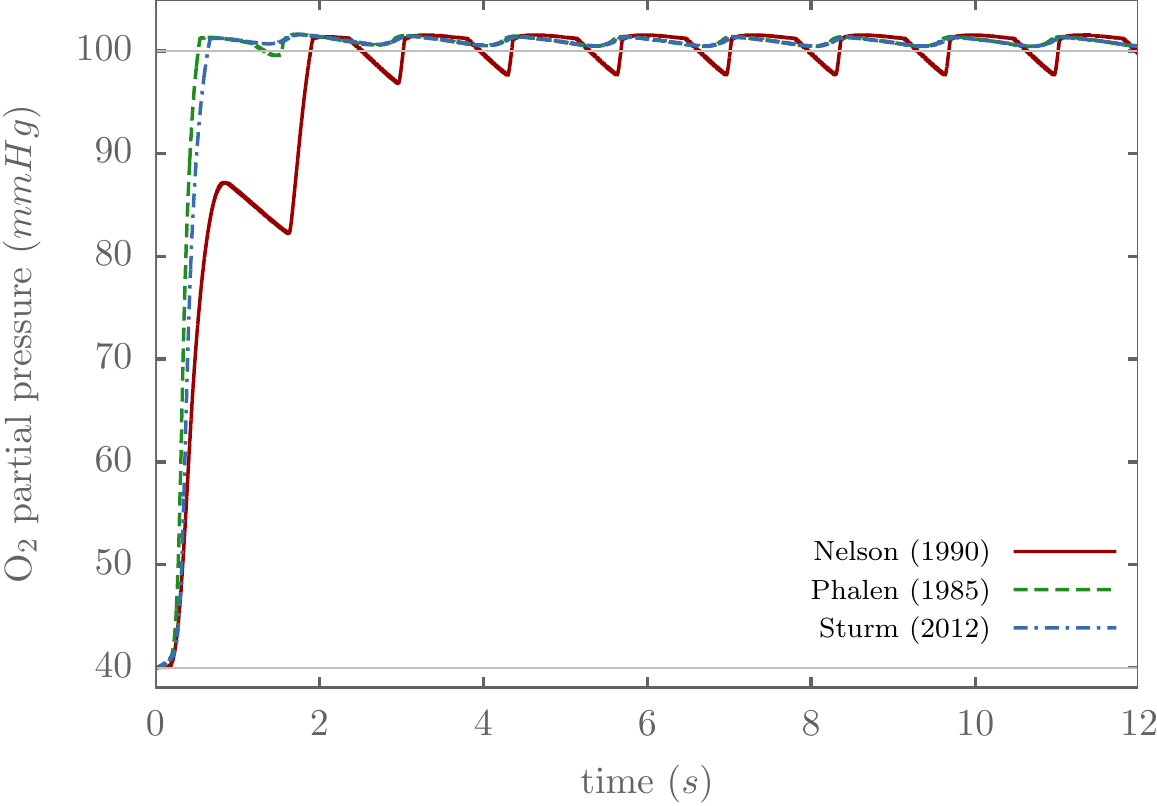}
  \includegraphics[width=0.49\textwidth]{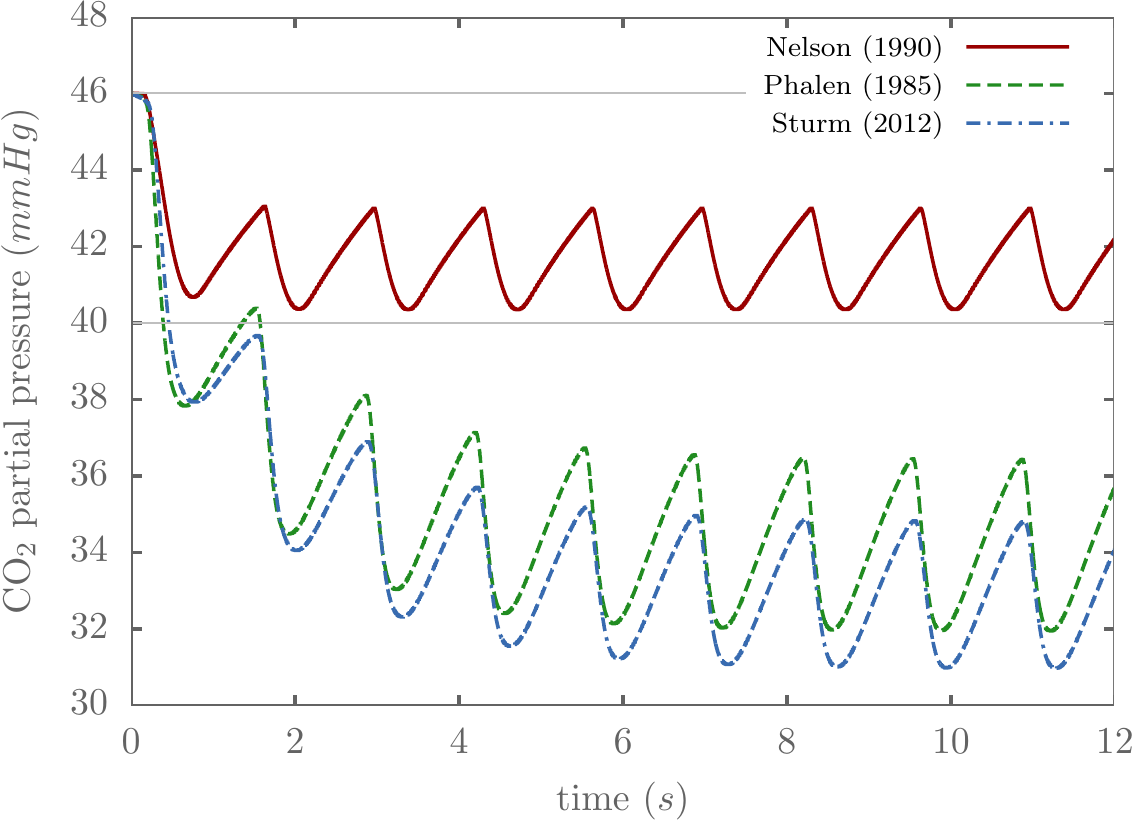}
\caption{Temporal variation of the mean partial pressure of oxygen (left) and carbon dioxide (right) in blood capillaries of adults (top) and infants (bottom), as predicted with our model for two bronchial tree representations.}
\label{FigO2CO2BloodAdultsInfants}
\end{figure}

Figure~\ref{FigO2CO2BloodAdultsInfants} exhibits temporal variations (over 10 breathing cycles) of the mean $P_{\text{O}_2}$ and $P_{\text{CO}_2}$  in the blood for adults and infants, respectively. Oxygen is quicker to reach saturation in the blood than carbon dioxide is. That is because while O$_2$ is stored only in the blood hemoglobin, CO$_2$ can be stored in the blood in different ways. Since the solubility of CO$_2$ is much higher than that of O$_2$, carbon dioxide is transported easier in the dissolved state. More than 60$\%$ of CO$_2$ is stored in the bicarbonate ionic form \cite{Rodney2013}, with 30$\%$ being stored in the hemoglobin and 7$\%$ in the plasma. 

Fgure~\ref{FigO2CO2BloodAdultsInfants} shows that the predictions of the mean $P_{\text{O}_2}$ and $P_{\text{CO}_2}$ based on the airways geometry~\cite{nelson1990fractal} have a fluctuation amplitude that is higher than those based on all the other 24-generation models. This is attributed to the overestimated lengths of the generations 14--25 (see Figure~\ref{figureAWLA} or Table~\ref{Table:models2}). Increasing the lengths $L_i$ of the generations where diffusion is dominant results in a smaller mass fluxes at the alveolar inlet, since $\partial_x C(x,t) \sim 1/ \sum L_i$. This damping effect results in smaller fluctuation amplitudes and in overall reduction in gas exchange. 

The model predictions of $P_{\text{O}_2}$ and $P_{\text{CO}_2}$ in Figs.~\ref{FigO2CO2BAAdults} and~\ref{FigO2CO2BloodAdultsInfants} demonstrate their sensitivity to the selection of a bronchial tree conceptualizations. Consequently, we use our model to ascertain the veracity of these alternative tree representations (Models A--H in Tables~\ref{Table:models2} and~\ref{Table:models3}). This is done by comparing the model predictions of key physiological quantities (total resistance, dead space volume, total path length, and min and max mean $P_{\text{O}_2}$ and $P_{\text{CO}_2}$ in the capillaries) with their observed values reported in the literature. For each airway geometry, we computed $P_{\text{O}_2}$ and $P_{\text{CO}_2}$ distributions along the capillaries every one eighth of a period over a whole period in the quasi-stationary state. In most cases, the quasi-stationary state was reached after ten breathing cycles. 
The minimum and maximum values of $P_{\text{O}_2}$ and $P_{\text{CO}_2}$ are obtained by averaging these distributions over the length of the capillary.
 The results are summarized in Table~\ref{Table:assessmentTable} together with the number of generations, and length and diameter of the terminals branches. The measured values of the dead space DS for adults and infants are taken as averages of those reported in \cite{Fowler1948, Biratg1959Respiratory, harris1973prediction, Lewis1979, Williams1997, tang2006new} and \cite{kerr1976dead, Epstein1980, Tilo1984, SANDBERG1987, Lagneaux1988, numa1996anatomic, Wenzel1999, Theodore2017}, respectively. The measured values of the resistance $\mathcal R$ for  adults and infants are taken as averages of those reported in~\cite{DuBOIS1955, bachofen1968, viljanen1982body, hantos1986, Kaczka1997, Koch2013, Sylvia2015} and \cite{Tilo1984, Wenzel1999, cook1957studies, swyer1960ventilation, Poglar1961, karlberg1962respiratory, polgar1965nasal, burnard1965pulmonary, Poglar1966, wohl69, Radford1974, stocks1977new, Thomas2006, Battisti2012}, respectively.  
 

\begin{table}[htbp]
\caption{Key physiological characteristics (dead space, DS; resistance, $\cal R$; total path length $L$; time needed to reach saturation of oxygen in blood, $T_r$; and mean partial pressures, $\bar P_{\text{O}_2}$ and $\bar P_{\text{CO}_2}$) computed with our model for alternative airways geometries of adults and infants. These predicted values are compared with their independently measured counterparts (M). $\dagger$ corresponds to spatial averages along the capillary by assuming that $P_{\text{O}_2}$ increases from 40 to 100 and $P_{\text{CO}_2}$ decreases from 46 to 40 over one third of the capillary length. Partial pressures are in units of cmH$_2$O.}
\centering
\begin{tabular}{lcccccccc}
\toprule
\multirow{2}[3]{*}{\scriptsize{Airways}} & DS & $\cal R$  & $L$ &  $T_r$ &  \multicolumn{2}{c}{$\bar P_{\text{O}_2}$}&\multicolumn{2}{c}{$\bar P_{\text{CO}_2}$} \\
\cmidrule(lr){6-7}\cmidrule(lr){8-9}
\scriptsize{geometry} & \scriptsize{(mL)} & \scriptsize{(cmH$_2$O.s/L)} & \scriptsize{(cm)} & \scriptsize{(sec)} & \scriptsize{(min)} & \scriptsize{(max)}  & \scriptsize{(min)} & \scriptsize{(max)}  \\
\midrule
\multicolumn{9}{c}{adults} \\
\cite{nelson1990fractal}	&	2681.62	&	1.49	&	29.46	&	 6	&	100.7	&	100.9	&	27	&	28.9	\\
\cite{weibel1965morphometry}	&	1656	&	0.184	&	27.28	&	 1.3	&	100.4	&	100.8	&	17.3	&	22.4	\\
\cite{ionescu2010viscoelasticity}	&	77537	&	0.037	&	33.93	&	 -	&	-	&	-	&	-	&	-	\\
\cite{horsfield1968morphology}	&	103141	&	0.037	&	34.53	&  -	&	-	&	-	&	-	&	-		\\
\cite{phalen1985postnatal}	&	71.07	&	0.357	&	19.4	& 	21.4 	&	96.2	&	101	&	42.9	&	45.4	\\
\cite{wiggs1990model}	&	830	&	0.56	&	27.27	&	 1.3	&	100.5	&	100.9	&	21.8	&	28.7	\\
\cite{Olson1970}	&	1759	&	0.049	&	31.17	&	 1.2	&	100.6	&	100.8	&	23.7	&	27.5	\\
M & 127-190 & 1.08-2.6 & 32 & - & 90$^\dagger$ & 90$^\dagger$ & 41$^\dagger$ & 41$^\dagger$ \\
\midrule
\multicolumn{9}{c}{infants} \\
\cite{nelson1990fractal} &	1.48	&	98.8	&	13.5	&	 2	&	97.6	&	101.5	&	40.3	&	43	\\
\cite{phalen1985postnatal}	&	16.58	&	2.06	&	15.06	&	 0.52	&	100.4	&	101.3	&	32	&	36.4		\\
\cite{sturm2012theoretical}	&	22.75	&	6.09	&	15.06	&  0.64	&	100.4	&	101.2	&	31	&	34.7	\\
M &  4.15-6.7&  22.6-52 &  &  & 90$^\dagger$ & 90$^\dagger$ & 41$^\dagger$ & 41$^\dagger$ \\
\bottomrule
\end{tabular}
  \label{Table:assessmentTable}
\end{table}

\begin{table}[htbp]
\caption{Computed average air flow-rate ($V$) entering the alveoli, blood flow-rate ($Q$) and ventilation to perfusion ratio($V/Q$) for the different airways geometries. These predicted values are compared with measurements reported in the literature (M). }
\centering
\begin{tabular}{lccc}
\toprule
 Airways & $V$& $Q$  & $V/Q$ \\
geometry & mm$^3$/s ($l/min$) & mm$^3$/s ($l/min$) &  \\
\midrule
\multicolumn{4}{c}{adults} \\
\cite{nelson1990fractal}	&	0.00065 (19.46)	&	0.00037 (11.28)	&	1.75	\\
\cite{weibel1965morphometry}	&	0.0007 (20.96)	&	0.00037 (11.28)		&	1.86	\\
\cite{phalen1985postnatal}	&	0.0007 (20.96)	&	0.00037 (11.28)		&	1.86	\\
\cite{wiggs1990model} 	&	0.00069 (20.66)	&	0.00037 (11.28)		&	1.86	\\
\cite{Olson1970} 	&	0.00069 (20.66)	&	0.00037 (11.28)		&	1.86	\\
M \cite{bryan1964factors} &  &  & 0.67-1.7 \\
\midrule
\multicolumn{4}{c}{infants} \\
\cite{nelson1990fractal}	&	0.00044 (0.40)	&	0.000057 (0.52)	&	0.77	\\
\cite{phalen1985postnatal}	&	0.000318 (2.9)	&	0.000057 (0.52)	&	5.58	\\
\cite{sturm2012theoretical}	&	0.000311 (2.84)	&	0.000057 (0.52)	&	5.46	\\
M\cite{dassios2017ventilation} &  &   &0.8-1.1  \\
\bottomrule
\end{tabular}
  \label{Table:assessmentVoQ}
\end{table}

The values of the total path length, $L$, total resistance, $\mathcal R$, and dead space DC in Table~\ref{Table:assessmentTable} for various bronchial tree geometries are estimated as 
\begin{align}
L = \sum_{k=0}^{N_\text{gen}}  {L_{k}}, \quad
{\cal R} = \sum_{k=0}^{N_\text{gen}} \frac{128 \mu L_k}{\pi D_k^4 2^k}, \quad
\text{DS} = \sum_{k=0}^{N_\text{gen}} 2^k \frac{\pi D_k^2  L_k}{4}
\label{eqLRDS}
\end{align}
where $\mu$ is the viscosity of air. In Table~\ref{Table:assessmentTable}, we compare thus computed values of dead space DS with their measured counterparts (see also Fig.~\ref{fig:FRCV_DS}); the latter are used in $N_\text{al} V_\text{al}(t=0) = \text{FRC} - \text{DS}$ to estimate the total alveolar volume. The expression for total resistance ${\cal R}$ is based on the assumption of Poiseuille flow in each channel. Due to losses at the branching junctions, the actual resistance is expected to be larger than ${\cal R}$, especially during expiration because the mixing loss associated with two channel flows merging into one is larger than that of a single channel flow splitting into two.

Table \ref{Table:assessmentTable} reveals that the resistance $\mathcal R$ estimated for the various airways geometries for adults is well below the range of reported measurements, except for the fractal model~\cite{nelson1990fractal} that results in the resistance that falls within the observed range. In general, the smaller the airways resistance, the sooner the mean partial pressures of oxygen and carbon dioxide reach their quasi-stationary state, i.e., the smaller $T_r$ (see, also, Fig.~\ref{FigO2CO2BloodAdultsInfants}). The only exception is the model~\cite{phalen1985postnatal}, whose estimate of the resistance $\mathcal R$ is well below the range of reported measurements but estimate of $T_r$ is an order of magnitude larger than the estimates derived from the other models. 

In infants, the resistance $\mathcal R$ predicted with the airways geometry~\cite{nelson1990fractal} is well above the observed range, while the estimates based on the geometries~\cite{phalen1985postnatal} and~\cite{sturm2012theoretical} are significantly below. As a consequence, airways geometries~\cite{phalen1985postnatal} and~\cite{sturm2012theoretical} predict $T_r < 1$~s, while the geometry~\cite{nelson1990fractal} results in $T_r > 2$~s. Both estimates are almost two orders of magnitude smaller than the observed value. The predictions of quasi-steady state $P_{\text{CO}_2}$ in the blood, based on the tree models~\cite{phalen1985postnatal} and~\cite{sturm2012theoretical}, oscillate in the range appreciably below their nominal range (40--46~mmHg) (Fig.~\ref{FigO2CO2BloodAdultsInfants}). As noted in~\cite{phalen1985postnatal}, abnormally elevated CO$_2$ levels in the blood are typically caused by hypoventilation, whereas reduced levels are  caused by hyperventilation. This finding enables us to categorize the airways geometries in Table~\ref{Table:assessmentTable} as either hyperventilating or hypoventilating.

In assessing the adults (infant) airways geometries, the blood flow-rate in the capillaries, $Q$, was set to its physiological value (Table \ref{tableModelParameters2}). The mean ventilation flow-rate, $V$, was computed as the ratio of the intake volume of air in one breathing cycle to the period of the cycle at the quasi-stationary state. Mechanical models $\dot{V}_\text{lu} = \dot{V}_\text{lu}(\Delta p)$, including the one in Eq.~\eqref{eqMECHMODEL}, suggest that $V$ is primarily a function of the pressure signal, compliance of the lung tissue and resistance of the airways. In assessing the alternative airways geometries, the same compliance and pressure signal was used for all the reported adult (infant) geometries, while the airways resistance varied with the airways geometry according to Eq.~\eqref{eqLRDS}. This strategy yields a one to one correspondence between the resistance (Table~\ref{Table:assessmentTable}) and the ventilation to perfusion ratio (Table~\ref{Table:assessmentVoQ}).

Table~\ref{Table:assessmentTable} also reveals that a higher dead space volume does not always lead to a smaller airways resistance, and vice versa: compare, e.g., the predictions based on the tree models \cite{phalen1985postnatal} and \cite{sturm2012theoretical} or \cite{nelson1990fractal} and  \cite{weibel1965morphometry}. For a given circular channel of length $L$ and diameter $D$, both the channel volume and the resistance would increase if $L$ increases by a factor $\alpha>1$ and $D$ increases by a factor $\beta$ that satisfies the condition $\alpha^{-1/2} < \beta < \alpha^{1/4}$.

Impact of the lung resistance and compliance on gas exchange with the blood is captured through the time scale of the mechanical model, $\tau = \cal{R C}$. This characteristic time, listed in Table \ref{Table:assessmentTableAUX} for the alternative airways geometries, characterizes how fast the system responds to changes in the driving pressure difference in term of the air volume exchange. The compliance $\mathcal C$, a property of the lung tissue, is taken to be the same in all the geometric models for adults (infants), since these differ only in terms of airways geometry. That is in contrast to the resistance $\mathcal R$, which  depends on the airways geometry. This implies that the resistance is a sufficient metric to compare the speed of volume exchange for the various adults (or infants) cases.

\begin{table}
\caption{Model predictions of characteristic time $\tau = \mathcal C \mathcal R$; the number of airways generations, $N_\text{gen}$; length and diameter of the last generation channels, $L_{N_\text{gen}}$ and $D_{N_\text{gen}}$; the ratio of the airways' inlets diameter per alveolus to the alveolar diameter, $R_\text{in} / R_\text{al}$; the Reynolds number in the last generation, $\mathtt{Re}_{N_\text{gen}} = \rho \bar u_{N_\text{gen}} D_{N_\text{gen}} / \mu$; and the P\'{e}clet number in the last generation, $\mathtt{Pe}_{N_\text{gen}} = L_{N_\text{gen}} \bar u_{N_\text{gen}} / \cal D$. Here $\rho$ and $\mu$ are the density viscosity of blood, respectively; ${\cal D}$ is the diffusion coefficient of species in air; and the mean speed $\bar u_{N_\text{gen}} = (2TVf)/2^{N_\text{gen}-1}$ is estimated by assuming that the lung changes its volume by the tidal volume, $TV$, over half a breath period, $1/(2f)$; and $f$ is the breathing frequency.}
\centering
\begin{tabular}{lccccccc}
\toprule
                   & $\tau$  & $N_\text{gen}$ & $L_{N_\text{gen}}$ & $D_{N_\text{gen}}$ & $R_\text{in}/R_\text{al}$ & $\mathtt{Re}_{N_\text{gen}}$ & $\mathtt{Pe}_{N_\text{gen}}$ \\
\scriptsize{Model} & \scriptsize{(sec)} & & \scriptsize{(mm)} & \scriptsize{(mm)} & & &  \\
\midrule
\multicolumn{8}{c}{adults} \\
\cite{nelson1990fractal}	&	0.14 &	24 & 1.49 & 0.328 &  0.045 & 0.01 & 0.04 	\\
\cite{weibel1965morphometry}	&	0.017 &	24 & 0.501 & 0.418 &  0.073	& 0.01 & 0.01 \\
\cite{ionescu2010viscoelasticity}	&	0.003 &	24 & 3.1 & 0.96 &  0.387	& & \\
\cite{horsfield1968morphology}	&	 0.003 & 25 & 	3.9 & 0.79 & 0.525 & & \\
\cite{phalen1985postnatal}	&	 0.033 & 16 & 1.55 & 0.454 &	0.0003 & 1.81	&	5.82	\\
\cite{wiggs1990model}	&	 0.052 &	24 & 0.501 & 0.29 & 0.035 & 0.01 & 0.02	\\
\cite{Olson1970}	&	 0.005 &	24 & 0.6 & 0.35 &  0.051 & 0.01 & 0.01	\\
\midrule
\multicolumn{8}{c}{infants} \\
\cite{nelson1990fractal}	&	0.43 &	15 & 0.87 & 0.13 &  0.0011	&	0.54	&	3.15	\\
\cite{phalen1985postnatal}	&	0.009 &	16 & 0.429 & 0.3 &  0.012	& 0.28	&	0.76	\\
\cite{sturm2012theoretical}	&	0.026 &	16 & 0.429 & 0.507 &  0.035	&	0.23	&	0.53 \\
\bottomrule
\end{tabular}
  \label{Table:assessmentTableAUX}
\end{table}

Since both length and diameter of the airways channels decrease with the generation number $n$, air flow and gases transport become increasingly dominated by diffusion, approaching the Stokes flow regime in the last generation, $n = N_\text{gen}$, as indicated by the low values of Reynolds ($\mathtt{Re}_{N_\text{gen}}$) and P\'eclet ($\mathtt{Pe}_{N_\text{gen}}$) numbers in Table~\ref{Table:assessmentTableAUX}. This implies that the diffusion term $A_\text{in} J_\text{diff}$ in Eq.~\ref{EQSPECIESALV} plays a significant role in the transfer of O$_2$ and CO$_2$ into and out of the alveolar space. The magnitude of this term depends on two quantities, which are related directly to the tree geometry: the concentration gradient, $\partial_x C(x_\text{in},t)$, and the area of the airways inlet to the alveolus, $A_\text{in}$. The former is affected by the spatial distribution of the species along the airways, while the latter is estimated as $A_\text{in}= 2^{N_\text{gen}-1}\pi D_{N_\text{gen}}^2 / (4 N_\text{al})$. Among the cases presented in Fig.~\ref{FigO2CO2BloodAdultsInfants}, the airways geometry~\cite{weibel1965morphometry} results in the lowest CO$_2$ levels and the highest $A_\text{in} = \pi R_\text{in}^2$ and $R_\text{in}/R_\text{al}$ (Table~\ref{Table:assessmentTableAUX}). The airways geometry~\cite{phalen1985postnatal} yields the highest CO$_2$ levels and the lowest $A_\text{in}$ and $R_\text{in}/R_\text{al}$. The difference in CO$_2$ removal from the blood, predicted with the tree models~\cite{weibel1965morphometry} and \cite{phalen1985postnatal}, is due to the fact the former consists of 24 generations while the latter comprises 16 generations only, with both sharing nearly similar values of the terminal branch diameter. The impact of the small alveolar inlet in the tree model~\cite{phalen1985postnatal} is also seen in Fig~\ref{FigO2CO2BloodAdultsInfants}, wherein the small flux at the alveolar inlet causes the oxygenation to be very slow, so that it takes 22~s (6 respiration cycles) for the blood to reach saturation.

\section {Discussion}

Quantitative description of the tracheobronchial airway morphometry is essential for the investigation of toxicologic effects of inhaled aerosols and gases under a variety of conditions. 
Limitations in measurements prohibit the accurate estimation of the dimensions of the generations beyond the first few. As a result, some studies opted to include only the first few generations in their design, while others relied on inferred dimensions for the rest of the generations, as discussed in the Introduction and Materials and Methods sections. Inaccurate measurements and corresponding quantitative descriptions of the airway anatomy undermine the ability of the associated dosimetry models in animal toxicology to accurately predict  particle transport and deposition dynamics within the respiratory tract.

A major goal of this study was to assess the adequacy of alternative lung airways geometries in terms of their ability to predict both observed geometric characteristics and observed lung function. Such an assessment is important because accurate predictions of the exchange of inhaled aerosols and gases with the blood in the pulmonary capillaries are not possible without a physically sound representation of the airways geometry. Such predictions require both an adequate pulmonary tree representation and a model to predict the dynamics of air and gases transport all the way down to the blood circulating in the pulmonary capillaries. A properly constructed mathematical representation of these processes not only enhances our understanding of pulmonary diseases, but also enables investigation of many what-if scenarios. 

To assess the veracity of alternative airways geometries in terms of their impact on predictions of the exchange of O$_2$ and CO$_2$ with the blood, we developed a spatially distributed model of unsteady transport of these gases throughout the bronchial tree. This model advances the  state of the art of distributed models by accounting for the coupling between O$_2$ and CO$_2$ (the oxyhemoglobin dissociation), and transport of these gases in the alveoli, across the alveolar membrane, and in the blood circulating the capillaries.  In our model, gas exchange with the blood is not only controlled by the ventilation to perfusion ratio (which is a purely mechanistic indicator), but is also limited by the conditions in the blood as constrained by the blood chemistry (dissociation curves), by diffusion-limited transport both in the small vessels (characterized by small Reynolds and P\'eclet numbers) and at the entrance to the alveoli, and by the gases concentrations at the inlets to the trachea and the capillaries (boundary conditions). The utility of our model extends beyond assessment of \emph{in vivo} networks to analysis and design of \emph{in silico} networks.

We posit that a physically sound representation of the airways geometry should meet topological, mechanical, and gas transport observables.   Any geometric representation of the tracheobronchial airway morphometry should conform to high-fidelity observable geometric quantities, such as the dimensions of the first few generations, dead space volume, and path length. Such a representation should also reproduce high-fidelity observables that characterize the mechanical behavior such as resistance and inductance. The associated lung mechanics model must also reproduce high-fidelity observables, including the tidal volume under normal breathing conditions. The degree to which alternative airways geometries reproduce these observables form a basis of a multi-faceted assessment tool to discriminate between them. 
Specifically, a proper representation of the airways geometry must ensure that the predicted dynamics of oxygenation and carbon dioxide removal in the blood agrees with physiologically significant observables, such as time to reach oxygen saturation in blood and mean partial pressure of O$_2$ and CO$_2$. We found that overestimation of the lengths of the bronchial generations wherein diffusion is dominant underpredicts mass fluxes at the alveolar inlet. This damping effect results in smaller fluctuation amplitudes and in overall reduction in gas exchange.  We also found  the area of the airways inlet to alveolus (terminal bronchiole diameter) to play a key role in predictions of the gases exchange. Thus extra care must be taken when reporting these dimensions. 

A suitable computational and/or experimental model that employs a physically sound representation of the airways dimensions can be ultimately used to design patient customized ventilation modalities, where the model is periodically updated with in vivo observations. We should not, however, ignore the fact the computational and/or experimental methods that incorporate the airways geometry used should be sufficiently accurate in a  way that is compatible with the objectives and scope of a particular study.

\section*{Conflict of Interest Statement}

The authors have no conflict of interest.

\section*{Acknowledgements} 
IL was supported by the American University of Beirut Research Board under award number 103371. LI funded by the Lebanese American University Research Fund under award number SRDC-r-2017-21. DMT was supported in part by National Science Foundation under award number DMS-1802189.

\bibliographystyle{plain}
\bibliography{airways-geometry}

\begin{thebibliography}{10}

\bibitem{bachofen1968}
H.~Bachofen.
\newblock Lung tissue resistance and pulmonary hysteresis.
\newblock {\em J. Appl. Physiol.}, 24(3):296--301, 1968.

\bibitem{Battisti2012}
O.~Battisti, J.~M. Bertrand, H.~Rouatbi, and G.~Escandar.
\newblock Lung compliance and airways resistance in healthy neonates.
\newblock {\em Pedatr. Therapeut.}, 2:1000114, 2012.

\bibitem{ben2006simplified}
Alona Ben-Tal.
\newblock Simplified models for gas exchange in the human lungs.
\newblock {\em J. Theor. Biol.}, 238(2):474--495, 2006.

\bibitem{Biratg1959Respiratory}
G.~Birath.
\newblock Respiratory dead space measurements in a model lung and healthy human
  subjects according to the single breath method.
\newblock {\em J. Appl. Physiol.}, 14(4):517--520, July 1959.

\bibitem{bryan1964factors}
A~C Bryan, L~G Bentivoglio, F~Beerel, H~MacLeish, A~Zidulka, and D~V Bates.
\newblock Factors affecting regional distribution of ventilation and perfusion
  in the lung.
\newblock {\em J. Appl. Physiol.}, 19(3):395--402, 1964.

\bibitem{burnard1965pulmonary}
E.~D. Burnard, P.~Grattan-Smith, C.~G. Picton-Warlow, and A.~Grauaug.
\newblock Pulmonary insufficiency in prematurity.
\newblock {\em J. Paediatr. Child Health}, 1(1):12--38, 1965.

\bibitem{Calay-2002-numerical}
R.~K. Calay, J.~Kurujareon, and A.~E. Hold\o.
\newblock Numerical simulation of respiratory flow patterns within human lung.
\newblock {\em Respir. Physiol. Neurobiol.}, 130(2):201--221, 2002.

\bibitem{carvalho2011respiratory}
Alysson~Roncally Carvalho and Walter~Araujo Zin.
\newblock Respiratory system dynamical mechanical properties: modeling in time
  and frequency domain.
\newblock {\em Biophys. Rev.}, 3(2):71, 2011.

\bibitem{cook1957studies}
C.~D. Cook, J.~M. Sutherland, S.~Segal, R.~B. Cherry, J.~Mead, M.~B. McIlroy,
  and C.~A. Smith.
\newblock Studies of respiratory physiology in the newborn infant. {III.
  M}easurements of mechanics of respiration.
\newblock {\em J. Clin. Investig.}, 36(3):440--448, 1957.

\bibitem{Theodore2017}
T.~Dassios, P.~Dixon, A.~Hickey, S.~Fouzas, and A.~Greenough.
\newblock Physiological and anatomical dead space in mechanically ventilated
  newborn infants.
\newblock {\em Pediatr. Pulmonol.}, 53(1):57--63, 2018.

\bibitem{dassios2017ventilation}
Theodore Dassios, Kamal Ali, Thomas Rossor, and Anne Greenough.
\newblock Ventilation/perfusion ratio and right to left shunt in healthy
  newborn infants.
\newblock {\em J. Clin. Monit. Comput.}, 31(6):1229--1234, 2017.

\bibitem{davidson1974transport}
M~R Davidson and J~M Fitz-Gerald.
\newblock Transport of {O}2 along a model pathway through the respiratory
  region of the lung.
\newblock {\em Bull. Math. Biol.}, 36:275--303, 1974.

\bibitem{desplechain1983pores}
C~Desplechain, B~Foliguet, E~Barrat, G~Grignon, and F~Touati.
\newblock The pores of kohn in pulmonary alveoli.
\newblock {\em Bulletin europeen de physiopathologie respiratoire},
  19(1):59--68, 1983.

\bibitem{DuBOIS1955}
A.~B. Dubois, S.~Y. Botelho, and Jr~Comroe, J.~H.
\newblock A new method for measuring airway resistance in man using a body
  plethysmograph: values in normal subjects and in patients with respiratory
  disease.
\newblock {\em J. Clin. Invest.}, 35(3):327--335, 1956.

\bibitem{Epstein1980}
R.~A. Epstein and A.~I. Hyman.
\newblock Ventilatory requirements of critically ill neonates.
\newblock {\em Anesthesiology}, 53(5):379--384, 1980.

\bibitem{Miller-1993-Lower}
J.~D.~Crapo F.~J.~Miller, R. R.~Mercer.
\newblock Lower respiratory tract structure of laboratory animals and humans:
  dosimetry implications.
\newblock {\em Aerosol. Sci. Technol.}, 18:257--271, 1993.

\bibitem{ferris1964partitioning}
B~G Ferris~Jr, Jere Mead, and L~H Opie.
\newblock Partitioning of respiratory flow resistance in man.
\newblock {\em J. Appl. Physiol.}, 19(4):653--658, 1964.

\bibitem{Fowler1948}
W.~S. Fowler.
\newblock Lung function studies. {II. The} respiratory dead space.
\newblock {\em Am. J. Physiol.}, 154(3):405--416, 1948.

\bibitem{Tilo1984}
T.~Gerhardt and E.~Bancalari.
\newblock Apnea of prematurity: {I. L}ung function and regulation of breathing.
\newblock {\em Pediatrics}, 74(1):58--62, 1984.

\bibitem{gheorghiu2005lung}
Stefan Gheorghiu, Signe Kjelstrup, Peter Pfeifer, and M-O Coppens.
\newblock Is the lung an optimal gas exchanger?
\newblock In G.~A. Losa, D.~Merlini, T.~F. Nonnenmacher, and E.~R. Weibel,
  editors, {\em Fractals in Biology and Medicine}, pages 31--42. Birkh\"auser
  Basel, 2005.

\bibitem{Guyton}
John~E Hall.
\newblock {\em Guyton and Hall Textbook of Medical Physiology: Enhanced
  E-book}.
\newblock Elsevier Health Sciences, 3 edition, 2010.

\bibitem{hantos1986}
Z.~Hantos, B.~Daroczy, B.~Suki, G.~Galgoczy, and T.~Csendes.
\newblock Forced oscillatory impedance of the respiratory system at low
  frequencies.
\newblock {\em J. Appl. Physiol.}, 60(1):123--132, 1986.

\bibitem{hardman2001respiratory}
J~G Hardman.
\newblock {\em Respiratory physiological modelling—the design, construction,
  validation and application of a set of original respiratory physiological
  models}.
\newblock PhD thesis, PhD thesis, Division of Anaesthesia and Intensive Care,
  University of Nottingham, 2001.

\bibitem{harper2001acoustic}
Paul Harper, Steve~S Kraman, Hans Pasterkamp, and George~R Wodicka.
\newblock An acoustic model of the respiratory tract.
\newblock {\em IEEE Trans. Biomed. Engrg.}, 48(5):543--550, 2001.

\bibitem{harris1973prediction}
E.~A. Harris, M.~E. Hunter, E.~R. Seelye, M.~Vedder, and R.~M.~L. Whitlock.
\newblock Prediction of the physiological dead-space in resting normal
  subjects.
\newblock {\em Clin. Sci.}, 45(3):375--386, 1973.

\bibitem{herring2014growth}
Matt~J Herring, Lei~F Putney, Gregory Wyatt, Walter~E Finkbeiner, and Dallas~M
  Hyde.
\newblock Growth of alveoli during postnatal development in humans based on
  stereological estimation.
\newblock {\em Am. J. Physiol. Lung Cell. Mol. Physiol.}, 307(4):L338--L344,
  2014.

\bibitem{horsfield1968morphology}
K.~Horsfield and G.~Cumming.
\newblock Morphology of the bronchial tree in man.
\newblock {\em J. Appl. Physiol.}, 24(3):373--83, 1968.

\bibitem{horsfield1971models}
K.~Horsfield, G.~Dart, D.~E. Olson, G.~F. Filley, and G.~Cumming.
\newblock Models of the human bronchial tree.
\newblock {\em J. Appl. Physiol.}, 31(2):207--217, 1971.

\bibitem{ionescu2010viscoelasticity}
C.-M. Ionescu, W.~Kosi{\'n}ski, and R.~De~Keyser.
\newblock Viscoelasticity and fractal structure in a model of human lungs.
\newblock {\em Arch. Mech.}, 62(1):21--48, 2010.

\bibitem{Kaczka1997}
D.~W. Kaczka, E.~P. Ingenito, B.~Suki, and K.~R. Lutchen.
\newblock Partitioning airway and lung tissue resistances in humans: effects of
  bronchoconstriction.
\newblock {\em J. Appl. Physiol.}, 82(5):1531--1541, 1997.

\bibitem{karlberg1962respiratory}
P.~Karlberg and G.~Koch.
\newblock Respiratory studies in newborn infants. {III.: D}evelopment of
  mechanics of breathing during the first week of life. {A} longitudinal study
  1.
\newblock {\em Acta Paediatr.}, 51:121--129, 1962.

\bibitem{kerr1976dead}
A.~A. Kerr.
\newblock Dead space ventilation in normal children and children with
  obstructive airways diease.
\newblock {\em Thorax}, 31(1):63--69, 1976.

\bibitem{Koch2013}
B.~Koch, N.~Friedrich, H.~V{\"o}lzke, R.~A. J{\"o}rres, S.~B. Felix, R.~Ewert,
  C.~Schaeper, and S.~Gl{\"a}ser.
\newblock Static lung volumes and airway resistance reference values in healthy
  adults.
\newblock {\em Respirology}, 18(1):170--178, 2013.

\bibitem{Lagneaux1988}
D.~Lagneaux, C.~Mossay, F.~Geubelle, and G.~Christiaens.
\newblock Alveolar data in healthy, awake neonates during spontaneous
  ventilation: a preliminary investigation.
\newblock {\em Pediatr. Pulmonol.}, 5(4):225--231, 1988.

\bibitem{lee2007fluid}
J.~W. Lee, M.~Y. Kang, H.~J. Yang, and E.~Lee.
\newblock Fluid-dynamic optimality in the generation-averaged
  length-to-diameter ratio of the human bronchial tree.
\newblock {\em Med. Biol. Engrg. Comput.}, 45(11):1071--1078, 2007.

\bibitem{Lewis1979}
S.~Lewis and C.~J. Martin.
\newblock Characteristics of the washout dead space.
\newblock {\em Resp. Physiol.}, 36(1):51--63, 1979.

\bibitem{liu98}
C.~H. Liu, S.~C. Niranjan, J.~W. Clark, K.~Y. San, J.~B. Zwischenberger,
  A.~Bidani, et~al.
\newblock Airway mechanics, gas exchange, and blood flow in a nonlinear model
  of the normal human lung.
\newblock {\em J. Appl. Physiol.}, 84(4):1447--1469, 1998.

\bibitem{martin2013modeling}
S.~Martin and B.~Maury.
\newblock Modeling of the oxygen transfer in the respiratory process.
\newblock {\em ESAIM Math. Model. Numer. Anal.}, 47(4):935--960, 2013.

\bibitem{mauroy20053d}
B~Mauroy.
\newblock 3d hydrodynamics in the upper human bronchial tree: Interplay between
  geometry and flow distribution.
\newblock In G.~A. Losa, D.~Merlini, T.~F. Nonnenmacher, and E.~R. Weibel,
  editors, {\em Fractals in Biology and Medicine}, pages 43--53. Birkh\"auser
  Basel, 2005.

\bibitem{mauroy2010influence}
B.~Mauroy and P.~Bokov.
\newblock The influence of variability on the optimal shape of an airway tree
  branching asymmetrically.
\newblock {\em Phys. Biol.}, 7(1):016007, 2010.

\bibitem{mead1969contribution}
J.~Mead.
\newblock Contribution of compliance of airways to frequency-dependent behavior
  of lungs.
\newblock {\em J. Appl. Physiol.}, 26(5):670--673, 1969.

\bibitem{Meade1972}
F.~Meade.
\newblock A formula for the carbon dioxide dissociation curve.
\newblock {\em Br. J. Anaesth.}, 44:630, 1972.

\bibitem{menache2008airway}
M.~G. M{\'e}nache, W.~Hofmann, B.~Ashgarian, and F.~J. Miller.
\newblock Airway geometry models of children's lungs for use in dosimetry
  modeling.
\newblock {\em Inhal. Toxicol.}, 20(2):101--126, 2008.

\bibitem{Montesantos-2016-creation}
S.~Montesantos, I.~Katz, M.~Pichelin, and G.~Caillibotte.
\newblock The creation and statistical evaluation of a deterministic model of
  the human bronchial tree from {HRCT} images.
\newblock {\em PLoS One}, 11(12):e0168026, 2016.

\bibitem{moukalled2016finite}
Fadl Moukalled, L~Mangani, Marwan Darwish, et~al.
\newblock {\em The finite volume method in computational fluid dynamics},
  volume 113.
\newblock Springer, 2016.

\bibitem{nelson1990fractal}
T.~R. Nelson, B.~J. West, and A.~L. Goldberger.
\newblock The fractal lung: universal and species-related scaling patterns.
\newblock {\em Experientia}, 46(3):251--254, 1990.

\bibitem{numa1996anatomic}
A.~H. Numa and C.~J. Newth.
\newblock Anatomic dead space in infants and children.
\newblock {\em J. Appl. Physiol.}, 80(5):1485--1489, 1996.

\bibitem{Ochs2004}
M.~Ochs, J.~R. Nyengaard, A.~Jung, L.~Knudsen, M.~Voigt, T.~Wahlers,
  J.~Richter, and H.~J.~G. Gundersen.
\newblock The number of alveoli in the human lung.
\newblock {\em Am. J. Respir. Crit. Care Med.}, 169(1):120--124, 2004.

\bibitem{Olson1970}
D.~E. Olson, G.~A. Dart, and G.~F. Filley.
\newblock Pressure drop and fluid flow regime of air inspired into the human
  lung.
\newblock {\em J. Appl. Physiol.}, 28(4):482--494, 1970.

\bibitem{paiva1984model}
Manuel Paiva and L~A Engel.
\newblock Model analysis of gas distribution within human lung acinus.
\newblock {\em J. Appl. Physiol.}, 56(2):418--425, 1984.

\bibitem{phalen1985postnatal}
R.~F. Phalen, M.~J. Oldham, C.~B. Beaucage, T.~T. Crocker, and J.~D. Mortensen.
\newblock Postnatal enlargement of human tracheobronchial airways and
  implications for particle deposition.
\newblock {\em Anatom. Rec.}, 212(4):368--380, 1985.

\bibitem{phalen1978application}
R.~F. Phalen, H.~C. Yeh, G.~M. Schum, and O.~G. Raabe.
\newblock Application of an idealized model to morphometry of the mammalian
  tracheobronchial tree.
\newblock {\em Anatom. Rec.}, 190(2):167--176, 1978.

\bibitem{Phillips-1997-on}
C.~G. Phillips and S.~R. Kaye.
\newblock On the asymmetry of bifurcations in the bronchial tree.
\newblock {\em Respir. Physiol.}, 107(1):85--98, 1997.

\bibitem{Poglar1961}
G.~Polgar.
\newblock Airway resistance in the newborn infant: preliminary communication.
\newblock {\em J. Pediatr.}, 59(6):915--921, 1961.

\bibitem{polgar1965nasal}
G.~Polgar and G.~P. Kong.
\newblock The nasal resistance of newborn infants.
\newblock {\em J. Pediatr.}, 67(4):557--567, 1965.

\bibitem{Poglar1966}
G.~Polgar and S.~T. String.
\newblock The viscous resistance of the lung tissues in newborn infants.
\newblock {\em J. Pediatr.}, 69(5):787--792, 1966.

\bibitem{raabe1976tracheobronchial}
O.~G. Raabe.
\newblock {\em Tracheobronchial geometry: human, dog, rat, hamster--a
  compilation of selected data from the project respiratory tract deposition
  models}.
\newblock US Energy Research and Development Administration, Division of
  Biomedical and Environmental Research, Washington, DC, 1976.

\bibitem{Radford1974}
M.~Radford.
\newblock Measurement of airway resistance and thoracic gas volume in infancy.
\newblock {\em Arch. Dis. Child.}, 49(8):611--615, 1974.

\bibitem{Rodney2013}
R.~A. Rhoades and D.~R. Bell.
\newblock {\em Medical phisiology: Principles for clinical medicine}.
\newblock Lippincott Williams \& Wilkins, 2013.

\bibitem{riley1951analysis}
RiL Riley and A~Cournand.
\newblock Analysis of factors affecting partial pressures of oxygen and carbon
  dioxide in gas and blood of lungs: theory.
\newblock {\em J. Appl. Physiol.}, 4(2):77--101, 1951.

\bibitem{rutledge1994ventsim}
Geoffrey~W Rutledge.
\newblock Ventsim: a simulation model of cardiopulmonary physiology.
\newblock In {\em Proceedings of the Annual Symposium on Computer Application
  in Medical Care}, page 878. American Medical Informatics Association, 1994.

\bibitem{rutledge1993design}
Geoffrey~W Rutledge, George~E Thomsen, Brad~R Farr, Maria~A Tovar, Jeanette~X
  Polaschek, Ingo~A Beinlich, Lewis~B Sheiner, and Lawrence~M Fagan.
\newblock The design and implementation of a ventilator-management advisor.
\newblock {\em Artif. Intell. Med.}, 5(1):67--82, 1993.

\bibitem{SANDBERG1987}
K.~Sandberg, B.~A. Sj{\"o}qvist, O.~Hjalmarson, and T.~Olsson.
\newblock Efficiency of ventilation in neonatal pulmonary maladaptation.
\newblock {\em Acta Paediatr.}, 76(1):30--36, 1987.

\bibitem{Sharan1989}
M.~Sharan, M.~P. Singh, and A.~Aminataei.
\newblock A mathematical model for the computation of the oxygen dissociation
  curve in human blood.
\newblock {\em BioSystems}, 22:249--260, 1989.

\bibitem{Smith-2018-Human}
B.~M. Smith, H.~Traboulsi, J.~H.~M. Austin, A.~Manichaikul, E.~A. Hoffman,
  E.~R. Bleecker, W.~V. Cardoso, C.~Cooper, D.~J. Couper, S.~M. Dashnaw,
  J.~Guo, M.~K. Han, N.~N. Hansel, E.~W. Hughes, D.~R.~Jacobs Jr, R.~E. Kanner,
  J.~D. Kaufman, E.~Kleerup, C.~L. Lin, K.~Liu, C.~M. {Lo Cascio}, F.~J.
  Martinez, J.~N. Nguyen, M.~R. Prince, S.~Rennard, S.~S. Rich, L.~Simon,
  Y.~Sun, Watson~K. E., P.~G. Woodruff, C.~J. Baglole, and R.~G. Barr.
\newblock Human airway branch variation and chronic obstructive pulmonary
  disease.
\newblock {\em Proc. Natl. Acad. Sci. U. S. A.}, 115(5):E974--E981, 2018.

\bibitem{stocks1977new}
J.~Stocks, N.~M. Levy, and S.~Godfrey.
\newblock A new apparatus for the accurate measurement of airway resistance in
  infancy.
\newblock {\em J. Appl. Physiol.}, 43(1):155--159, 1977.

\bibitem{strauss2000relative}
G~Strauss-Blasche, M~Moser, M~Voica, DR~McLeod, N~Klammer, and W~Marktl.
\newblock Relative timing of inspiration and expiration affects respiratory
  sinus arrhythmia.
\newblock {\em Clinical and Experimental Pharmacology and Physiology},
  27(8):601--606, 2000.

\bibitem{sturm2012theoretical}
R.~Sturm.
\newblock Theoretical models of carcinogenic particle deposition and clearance
  in children's lungs.
\newblock {\em J. Thorac. Dis.}, 4(4):368--376, 2012.

\bibitem{swan2010evidence}
Annalisa~J Swan and Merryn~H Tawhai.
\newblock Evidence for minimal oxygen heterogeneity in the healthy human
  pulmonary acinus.
\newblock {\em J. Appl. Physiol.}, 110(2):528--537, 2010.

\bibitem{swyer1960ventilation}
P.~R. Swyer, R.~C. Reiman, and J.~J. Wright.
\newblock Ventilation and ventilatory mechanics in the newborn: methods and
  results in 15 resting infants.
\newblock {\em J. Pediatr.}, 56(5):612--622, 1960.

\bibitem{tang2006new}
Y.~Tang, M.~J. Turner, and A.~B. Baker.
\newblock A new equal area method to calculate and represent physiologic,
  anatomical, and alveolar dead spaces.
\newblock {\em Anesthesiology}, 104(4):696--700, 2006.

\bibitem{Thomas2006}
M.~R. Thomas, G.~F. Rafferty, R.~Blowes, J.~L. Peacock, N.~Marlow, S.~Calvert,
  A.~Milner, and A.~Greenough.
\newblock Plethysmograph and interrupter resistance measurements in prematurely
  born young children.
\newblock {\em Arch. Dis. Child Fetal Neonatal Ed.}, 91(3):F193--F196, 2006.

\bibitem{Thompson-1942-Growth}
D.~W. Thompson.
\newblock {\em Growth and Form}.
\newblock MacMillan, New York, 1942.
\newblock pp.~948-957.

\bibitem{Thurlbeck1982}
W.~M. Thurlbeck.
\newblock Postnatal human lung growth.
\newblock {\em Thorax}, 37:564--571, 1982.

\bibitem{Sylvia2015}
S.~Verbanck, A.~{Van Muylem}, D.~Schuermans, I.~Bautmans, B.~Thompson, and
  W.~Vincken.
\newblock Transfer factor, lung volumes, resistance and ventilation
  distribution in healthy adults.
\newblock {\em Eur. Respir. J.}, 47:166--176, 2016.

\bibitem{viljanen1982body}
A.~A. Viljanen, B.~C. Viljanen, P.~K. Halttunen, and K.-E. Kreus.
\newblock Body plethysmographic studies in non-smoking, healthy adults.
\newblock {\em Scand. J. Clin. Lab. Invest.}, 42(sup159):35--50, 1982.

\bibitem{warren1991red}
G.~L. Warren, K.~J. Cureton, W.~F. Middendorf, C.~A. Ray, and J.~A. Warren.
\newblock Red blood cell pulmonary capillary transit time during exercise in
  athletes.
\newblock {\em Med. Sci. Sports Exerc.}, 23(12):1353, 1991.

\bibitem{weibel1963morphology}
E.~R. Weibel.
\newblock {\em Morphology of the human lung}.
\newblock Academic Press, New York, 1963.

\bibitem{weibel1965morphometry}
E.~R Weibel.
\newblock {\em Morphometry of the human lung}.
\newblock Springer, 1965.

\bibitem{weibel2014morphometry}
E.~R. Weibel, A.~F. Cournand, and D.~W. Richards.
\newblock {\em Morphometry of the human lung}.
\newblock Springer, 2014.

\bibitem{weibel08}
E.~R. Weibel and B.~W. Knight.
\newblock A morphometric study on the thickness of the pulmonary air-blood
  barrier.
\newblock {\em J. Cell Biol.}, 21(3):367--384, 1964.

\bibitem{weibel2005mandelbrot}
Ewald~R Weibel.
\newblock Mandelbrot’s fractals and the geometry of life: A tribute to
  {B}eno{\^\i}t {M}andelbrot on his 80th birthday.
\newblock In G.~A. Losa, D.~Merlini, T.~F. Nonnenmacher, and E.~R. Weibel,
  editors, {\em Fractals in Biology and Medicine}, pages 3--16. Birkh\"auser
  Basel, 2005.

\bibitem{Wenzel1999}
U.~Wenzel, R.~R. Wauer, and G.~Schmalisch.
\newblock Comparison of different methods for dead space measurements in
  ventilated newborns using {CO2}-volume plot.
\newblock {\em Intensive Care Med.}, 25(7):705--713, 1999.

\bibitem{west1986beyond}
B.~J. West, V.~Bhargava, and A.~L. Goldberger.
\newblock Beyond the principle of similitude: renormalization in the bronchial
  tree.
\newblock {\em J. Appl. Physiol.}, 60(3):1089--1097, 1986.

\bibitem{wiggs1990model}
B.~R. Wiggs, R.~Moreno, J.~C. Hogg, C.~Hilliam, and P.~D. Pare.
\newblock A model of the mechanics of airway narrowing.
\newblock {\em J. Appl. Physiol.}, 69(3):849--860, 1990.

\bibitem{Williams1997}
E.~M. Williams, R.~M. Hamilton, L.~Sutton, J.~P. Viale, and C.~E.~W. Hahn.
\newblock Alveolar and dead space volume measured by oscillations of inspired
  oxygen in awake adults.
\newblock {\em Am. J. Respir. Crit. Care Med.}, 156(6):1834--1839, 1997.

\bibitem{wohl69}
M.~E.~B. Wohl, L.~C. Stigol, and J.~Mead.
\newblock Resistance of the total respiratory system in healthy infants and
  infants with bronchiolitis.
\newblock {\em Pediatrics}, 43(4):495--509, 1969.

\bibitem{zimmerman-2018-hydrodynamic}
R.~A. Zimmerman, G.~Severino, and D.~M. Tartakovsky.
\newblock Hydrodynamic dispersion in a tube with diffusive losses through its
  walls.
\newblock {\em J. Fluid Mech.}, 837:546--561, 2018.

\end{thebibliography}

\newpage
\section*{Figure Legends}

\begin{description}
\item[Figure 1.] Dependence of the length (left column) and diameter (right column), normalized with the corresponding length and diameter of the trachea, on the generation order $n$, in human adults (top row) and infants (bottom row).

\item[Figure 2.] Model workflow showing connections between constituent sub-models, as well as data streams / model inputs.

\item[Figure 3.] Alternative circuit models of gas flow in the bronchial tree: a general model (a) and its successive simplifications (b--d).

\item[Figure 4.] Dependence of functional residual capacity FRC (left) and dead space DS (right) on body height, which serves as a proxy for age. The correlation between FRC and lung volume/age reported in~\cite{Thurlbeck1982} were used to convert total lung volume into FRC. When needed, growth charts were used to express height as function of age. [a]: Engstom et al.~(1956), [b]: Cook and Hamann~(1961), [c]:  Engstrom et al.~(1962), [d]: Hart et al.~(1963), [e]: Doershuk et al.~(1970), [f]: Wood et al.~(1971), [g]: Hatch and Taylor~(1976), [h]: Stocks and Godfrey~(1977), [i]:  Greenough et al.~(1986), [j]: Roberts et al.~(1991), [k]: Tepper et al.~(1993), [l]: Merth et al.~(1995), [m]: Stocks and Quanjer~(1995), [n]: McCoy et al.~(1995), [o]: Neder et al.~(1999), [p]: Castile et al.~(2000), [q]: Hulskamp et al.~(2003), [r]:  Hislop et al.~(1986), [s]: Angus and Thurlbeck~(1972), [t]:  Davies and Reid~(1970), [u]:  Auld et al.~(1963), [v]:  Hart et al.~(1963). $\star$: Stocks and Quanjer~(1995) using Tepper et al.~(1993); Hanrahan et al.~(1990); Merth et al.~(1995); Taussig et al.~(1977); Greenough et al.~(1986).

\item[Figure 5.] Temporal variation of the pressure signal $\Delta p(t)$ (the model's input) and the lung volume $V_\text{lu}(t)$ (the model's output) under normal breathing conditions for an adult.

\item[Figure 6.] Each terminal branch is connected to a number of alveoli. Our model assumes that the bronchiole is connected to the center of each compliant spherical alveolus.

\item[Figure 7.] Dependence of the number of alveoli, $N_\text{al}$, on age of adults (left) and infants (right). 

\item[Figure 8.] A schematic of the mesh used in the finite volume method for a branch in generation $i$ and the two daughter branches in generation $i+1$.

\item[Figure 9.] Temporal variation of the mean partial pressure of oxygen (left) and carbon dioxide (right) in the blood and in the alveolar space of adults, as predicted with our model for two bronchial tree representations. 

\item[Figure 10.] Temporal variation of the mean partial pressure of oxygen (left) and carbon dioxide (right) in blood capillaries of adults (top) and infants (bottom), as predicted with our model for two bronchial tree representations.

\end{description}

\end{document}